\documentclass[12pt]{article}

\title{Hamiltonian and Brownian systems with long-range interactions: V. Stochastic kinetic equations \\
and theory of fluctuations}

\usepackage{amsthm,amsmath,amssymb,a4wide,psfig,epsf,color}

cm \hsize=18 true cm \topmargin=-2cm \oddsidemargin=-0.5cm
\def\mb#1{\setbox0=\hbox{$#1$}\kern-.025em\copy0\kern-\wd0
\kern-0.05em\copy0\kern-\wd0\kern-.025em\raise.0233em\box0}

\begin{document}

\author{Pierre-Henri Chavanis}
\maketitle
\begin{center}
Laboratoire de Physique Th\'eorique (CNRS UMR 5152), \\
Universit\'e
Paul Sabatier,\\ 118, route de Narbonne, 31062 Toulouse Cedex 4, France\\
E-mail: {\it chavanis{@}irsamc.ups-tlse.fr\\
 }
\end{center}

\begin{abstract}

We develop a theory of fluctuations for Brownian systems with weak
long-range interactions. For these systems, there exists a critical
point separating a homogeneous phase from an inhomogeneous
phase. Starting from the stochastic Smoluchowski equation governing
the evolution of the fluctuating density field of the Brownian
particles, we determine the expression of the correlation function of
the density fluctuations around a spatially homogeneous equilibrium
distribution. In the stable regime, we find that the temporal
correlation function of the Fourier components of the density
fluctuations decays exponentially rapidly with the same rate as the
one characterizing the damping of a perturbation governed by the
deterministic mean field Smoluchowski equation (without noise). On the
other hand, the amplitude of the spatial correlation function in
Fourier space diverges at the critical point $T=T_{c}$ (or at the
instability threshold $k=k_{m}$) implying that the mean field
approximation breaks down close to the critical point and that the
phase transition from the homogeneous phase to the inhomogeneous phase
occurs sooner.  By contrast, the correlations of the velocity
fluctuations remain finite at the critical point (or at the
instability threshold). We give explicit examples for the Brownian
Mean Field (BMF) model and for Brownian particles interacting via the
gravitational potential and via the attractive Yukawa potential. We
also introduce a stochastic model of chemotaxis for bacterial
populations generalizing the deterministic mean field Keller-Segel
model by taking into account fluctuations and memory effects.

\end{abstract}

\maketitle

\section{Introduction}
\label{sec_introduction}

In a recent series of papers \cite{hb1,hb2,hb3,hb4,assise}, we have
considered some theoretical aspects of the dynamics and thermodynamics
of systems with weak long-range interactions
\cite{dauxois,assisetot}. In these systems, the interaction potential
$u(r)$ decays with a rate slower than $1/r^d$ at large distances,
where $d$ is the dimension of space (these potentials are sometimes
called ``non-integrable''). As a result, any particle feels a
potential dominated by interactions with far away particles (i.e. the
interaction is not restricted to nearest neighbours) and the energy is
{\it non-additive}. This can lead to striking properties (absent in
systems with short-range interactions) such as inequivalence of
statistical ensembles and negative specific heats in the
microcanonical ensemble. On the other hand, the usual thermodynamic
limit $N\rightarrow +\infty$ with $N/V$ fixed is not relevant for
these systems and must be reconsidered. If we write the potential of
interaction as $u(|{\bf r}-{\bf r}'|)=k\tilde{u}(|{\bf r}-{\bf r}'|)$
where $k$ is the coupling constant, then the appropriate thermodynamic
limit for weak long-range interactions corresponds to $N\rightarrow
+\infty$ in such a way that the coupling constant $k\sim
1/N\rightarrow 0$ and the volume $V\sim 1$. In that limit, we have an
{\it extensive} scaling of the energy $E\sim N$ and entropy $S\sim N$
(while the temperature $T\sim 1$), but the system remains
fundamentally {\it non-additive}. Other equivalent combinations of the
parameters are possible to define the thermodynamic limit as discussed
in
\cite{hb1,assise} and in various contributions of \cite{assisetot}.
For systems with weak long-range interactions, it is often claimed
that the mean field approximation is a very good approximation and
that it becomes exact in the proper thermodynamic limit $N\rightarrow
+\infty$. In fact, this is true only if we are far from a critical
point. Close to a critical point, the fluctuations become large and
cannot be ignored. In that case, the two-body correlation function
does not factor out in a product of two one-body distribution
functions and the mean field approximation breaks down.  It is
therefore highly desirable to derive stochastic kinetic equations that
go beyond the mean field approximation and that take full account of
fluctuations. This is the main object of the present paper.

In our previous studies, we have distinguished two types of systems:
Hamiltonian and Brownian. Hamiltonian systems with long-range
interactions are {\it isolated} and evolve at fixed energy. The
dynamics of the particles is described by $N$ coupled deterministic
Newton equations. Since the energy is conserved, the relevant
statistical ensemble is the microcanonical ensemble. The evolution of
the $N$-body distribution function is governed by the Liouville
equation and the statistical equilibrium state is described by the
microcanonical distribution. Examples of such systems are provided by
stellar systems
\cite{saslawbook,bt,hut,paddy,review}, two-dimensional 
vortices \cite{km,sommeria,tabeling,houches} and the Hamiltonian Mean
Field (HMF) model
\cite{ar,cvb}; see also the important list of references 
in these papers. Brownian systems with long-range interactions, on
the other hand, are {\it dissipative} and evolve at fixed
temperature. The particles are subject to their mutual long-range
interactions but they experience, in addition, a friction force and a
stochastic force which mimic the interaction with a thermal bath
(that is due to {\it short-range} interactions). Therefore, the
dynamics of the particles is described by $N$ coupled stochastic
Langevin equations. The temperature is defined through the Einstein
relation as the ratio between the diffusion coefficient and the
friction coefficient, and it measures the strength of the stochastic
force.  Since the temperature is fixed, the relevant statistical
ensemble is the canonical ensemble. The evolution of the $N$-body
distribution function is governed by the Fokker-Planck equation and
the statistical equilibrium state is described by the canonical
distribution. Examples of such systems are provided by
self-gravitating Brownian particles
\cite{crs,virial2},  bacterial populations experiencing chemotaxis 
\cite{ks,ng,bio,nfp} and the Brownian Mean Field (BMF) model \cite{cvb,bco}.

Systems with long-range interactions have a very peculiar dynamics and
thermodynamics \cite{dauxois,assisetot}. When the interaction is attractive,
there exists a critical point separating a spatially homogeneous
(gaseous) phase from a spatially inhomogeneous (clustered) phase. In
the microcanonical ensemble (MCE), the ``clustered'' phase appears
below a critical energy $E_{c}$ and in the canonical ensemble (CE), it
appears below a critical temperature $T_{c}$. The homogeneous phase
may still exist for $E<E_{c}$ or $T<T_{c}$ but is dynamically and
thermodynamically unstable. One fundamental illustration of this type
of phase transitions corresponds to the famous Jeans instability in
astrophysics \cite{jeans,bt}. For the Jeans problem \footnote{As is
well-known, the standard Jeans analysis assumes that an infinite and
homogeneous self-gravitating system is a steady state of the
hydrodynamical equations (the so-called Euler-Poisson system), which
is not correct. This inconsistency in the stability analysis is
referred to as the {\it Jeans swindle}
\cite{bt}. However, the Jeans treatment can be justified in cosmology
\cite{peebles} if we take into account the expansion of the universe
because this creates a sort of ``neutralizing background'' in the
comoving frame expanding with the system allowing for infinite and
homogeneous steady states.}, the critical temperature $T_{c}=+\infty$
so that the system is always in the ``clustered'' phase: this
corresponds to the universe that we know, filled of galaxies. A
spatially homogeneous distribution of matter is always unstable to
sufficiently large wavelengths, above the Jeans length
$k_{J}^{-1}=(4\pi Gn m^2\beta)^{-1/2}$, and this leads to
gravitational collapse and clustering.  If we formally consider a
screened Newtonian interaction (attractive Yukawa potential) with
screening length $k_{0}^{-1}$, we find that the phase transition
occurs at a finite critical temperature $T_{c}=4\pi Gnm^2/k_{0}^2$
separating a homogeneous phase from a clustered phase \cite{hb1}. For
$T>T_{c}$ the homogeneous phase is stable and for $T<T_{c}$ it becomes
unstable to wavenumbers $k<k_{m}\equiv
k_{0}(T_c/T-1)^{1/2}$. Alternatively, we can consider a spatially
inhomogeneous self-gravitating gas in a spherical box of radius
$R$. In that case, there exists a critical energy $E_{c}=-0.335GM^2/R$
in MCE (discovered by Antonov
\cite{antonov,lbw} for stellar systems) and a critical temperature
$T_{c}=GMm/(2.52k_B R)$ in CE (discovered by Emden \cite{emden} for
isothermal stars and recently emphasized by the author for molecular
clouds in contact with a thermal bath \cite{aaiso} and for
self-gravitating Brownian particles
\cite{crs}) below which the system passes from a slightly inhomogeneous 
gaseous phase to a highly  inhomogeneous condensed phase \footnote{In
order to have a well-defined condensed phase, one has to introduce a
small-scale regularization of the gravitational potential. More
fundamentally, we can invoke quantum mechanics (Pauli exclusion
principle) and consider the case of self-gravitating fermions
(e.g. white dwarfs, neutron stars and fermion balls)
\cite{review}.}. Interestingly, the analogue of these ``gravitational phase
transitions'' also takes place in the context of the chemotaxis of
bacterial populations in biology \cite{jeansbio,jeansbio2}. In these
systems, an infinite and uniform distribution of cells {\it is} a
steady state of the equations of motion (the so-called Keller-Segel model and
its generalizations) so there is no ``Jeans swindle''. Furthermore, the
screened Yukawa potential enters naturally in the problem and the
screening length has a clear physical interpretation as it takes into
account the degradation of the secreted chemical. Another illustration
of this type of phase transitions is given by the study of toy models
like the Hamiltonian Mean Field (HMF) and the Brownian Mean Field
(BMF) models
\cite{ar,cvb}. For these systems, there exists a critical energy
$E_{c}=kM^2/(8\pi)$ and a critical temperature $T_{c}=kM/(4\pi)$
separating a spatially homogeneous (non-magnetized) phase from a
spatially inhomogeneous (magnetized) phase. The magnetization plays
the role of the order parameter and the phase transition is second
order in that case
\cite{ar,cvb}. We expect the above-mentioned types of
phase transitions, and their generalizations
\cite{klein,ellis,martzel,bb,touchette,review,hb1}, to occur for a
large class of mean field systems with different potentials of
interaction.

The critical point separating the homogeneous phase from the
inhomogeneous phase is often called a {\it spinodal point}. The
instability threshold of the homogeneous phase can be obtained from
different classical methods: (i) by studying the sign of the second
order variations of the thermodynamical potential (entropy in MCE or
free energy in CE) and determining when the spatially uniform
distribution becomes an unstable saddle point (see, e.g., Sec. 4.4. of
Paper I or Appendix C of \cite{nfp}) (ii) by studying the bifurcation
(from homogeneous to inhomogeneous) of the solutions of the meanfield
integrodifferential equation (I-19) determining the statistical
equilibrium state (see, e.g., Sec. 2.3 of Paper I and Appendix C of
\cite{nfp}) (iii) by studying the linear and nonlinear dynamical
stability of the homogeneous state with respect to kinetic equations:
Vlasov, Euler, Landau, Kramers, Smoluchowski... (see, e.g., Paper II and 
\cite{jeansbio2,nfp}). In the
homogeneous phase, the one-body distribution function is trivial (it
is spatially uniform with a Maxwellian velocity distribution) and the
state of the system is usually characterized by the two-body
distribution function. For weak long-range interactions, the two-body
correlation function $h(|{\bf r}_{1}-{\bf r}_{2}|)$ can be obtained
from the equilibrium BBGKY-hierarchy at the order $O(1/N)$ by
neglecting the three-body correlation function that is of order
$O(1/N^2)$
\cite{hb1}. This is similar to the Debye-H\"uckel approximation in
plasma physics. It is found however that, for large but fixed $N$, the
Fourier transform of the correlation function {diverges} at the
critical point (or at the instability threshold) so that the
mean field approximation breaks down close to the critical point. This
implies that the phase transition should take place {\it strictly
before} the critical point (or strictly before the instability
threshold) predicted by mean field theory. In
\cite{hb1,cvb}, we have reached this conclusion from the study of the
equilibrium BBGKY hierarchy. In the present work, we would like to
complement our previous studies by developing a theory of fluctuations
starting directly from the stochastic kinetic equation governing
the evolution of the fluctuating density field. In this paper, we restrict
ourselves to the case of Brownian systems.

The paper is organized as follows. In Sec. \ref{sec_overdamped}, we
consider a gas of Brownian particles in interaction in an overdamped
limit where inertial effects are neglected. In Sec. \ref{sec_ea},
using a BBGKY-like hierarchy, we give the deterministic kinetic
equation (\ref{ea5}) satisfied by the smooth density profile and the
Smoluchowski equation (\ref{ea7}) resulting from a mean field
approximation. In Sec. \ref{sec_ex}, we give the stochastic kinetic
equation (\ref{ex2}) satisfied by the exact density distribution
(expressed in terms of $\delta$-functions) and, averaging over the
noise, we recover the equation obtained from the BBGKY-like
hierarchy. In Sec. \ref{sec_cg}, we give the stochastic kinetic
equation (\ref{cg2}) satisfied by the coarse-grained density
distribution obtained by averaging the exact density distribution over
a small spatio-temporal window (thus keeping track of fluctuations).
In Appendix \ref{sec_ll}, we provide another derivation of this
equation by using the general theory of fluctuations exposed by Landau
\& Lifshitz \cite{ll}.  In Sec. \ref{sec_ch}, we derive generalized
stochastic Cahn-Hilliard equations in the limit of short-range
interactions. In Sec. \ref{sec_f}, starting from the stochastic
Smoluchowski equation (\ref{f1})-(\ref{f2}), we develop a theory of
fluctuations for Brownian particles with weak long-range
interactions. Specifically, we study the correlations $\langle
\delta\hat{\rho}_{k}(t)\delta\hat{\rho}_{k'}(t+\tau)\rangle$ of the Fourier
transform of the density fluctuations and show that it behaves like
$A(k) e^{\sigma(k)\tau}$. In the stable regime, the decay rate
$\sigma(k)<0$ of the temporal correlations coincides with the decay
rate of the perturbations $\delta\hat{\rho}_{k}(t)\sim e^{\sigma(k)t}$
governed by the deterministic mean field Smoluchowski equation
(without noise). On the other hand, for $T<T_{c}$, the amplitude of
the correlation function $A(k)$ diverges as we approach the
instability threshold $k\rightarrow k_m$, suggesting that the
instability takes place sooner than predicted by mean field theory
(for $T>T_{c}$, considering the ``dangerous'' mode $k=k_{*}$, the
correlation function diverges as we approach the critical point
$T\rightarrow T_{c}^{+}$). These results clearly demonstrate that
fluctuations cannot be ignored close to a critical point. In
Sec. \ref{sec_inertial}, we generalize our results to the case of an
inertial model of particles in interaction including a friction force
and a stochastic force. The overdamped model is recovered in a strong
friction limit. Again, the correlations of the density fluctuations
diverge as we approach the instability threshold but, in sharp
contrast, the correlations of the velocity fluctuations remain finite
at this threshold. In Sec.  \ref{sec_mm}, we derive a generalized
Smoluchowski equation taking into account memory effects and make
contact with the Cattaneo model and the telegraph equation. In
Sec. \ref{sec_chemo}, we derive a stochastic model of chemotaxis
generalizing the deterministic mean field Keller-Segel model by
keeping track of fluctuations. Finally, in Sec. \ref{sec_ps}, we
extend our study in phase space taking full account of the inertia of
the particles. We derive a stochastic Kramers equation and make the
connection with the diffusive and hydrodynamic models of the previous
sections.

\section{The overdamped case}
\label{sec_overdamped}

\subsection{The smooth density distribution}
\label{sec_ea}

We consider an overdamped  system of $N$ Brownian particles in interaction whose dynamics is governed by the coupled stochastic equations (see Paper II):
\begin{equation}
\label{ea1}
\frac{d{\bf r}_{i}}{dt}=-\mu m^2\nabla_i U+\sqrt{2D_{*}}{\bf R}_{i}(t),
\end{equation}
where $\mu=1/(m\xi)$ is the mobility ($\xi$ denotes the friction
coefficient), $U({\bf r}_{1},...,{\bf r}_{N})=\sum_{i<j}u(|{\bf
r}_{i}-{\bf r}_{j}|)$ is the potential of interaction, $D_{*}$ is the
diffusion coefficient and ${\bf R}_{i}(t)$ is a white noise such that
$\langle {\bf R}_{i}(t)\rangle={\bf 0}$ and $\langle
R_{i}^{\alpha}(t)R_{j}^{\beta}(t')\rangle=\delta_{ij}\delta_{\alpha\beta}\delta(t-t')$
where $i=1,...,N$ refer to the particles and $\alpha=1,...,d$ to the
coordinates of space.  We assume that the particles interact via a
binary potential $u(|{\bf r}_{i}-{\bf r}_{j}|)$ depending only on the
absolute distance between the particles. As discussed in the
Introduction, this stochastic model can describe self-gravitating
Brownian particles \cite{crs,virial2}, bacterial populations experiencing
chemotaxis \cite{ng,bio,nfp} (see also Sec. \ref{sec_chemo}) or toy
models like the BMF model
\cite{cvb}. The $N$-body distribution $P_{N}({\bf r}_{1},...,{\bf
r}_{N},t)$ is solution of the Fokker-Planck equation
\begin{eqnarray}
\label{ea2}
\frac{\partial P_{N}}{\partial t}
=\sum_{i=1}^{N}\frac{\partial}{\partial {\bf r}_{i}}\cdot \left (D_{*}\frac{\partial P_{N}}{\partial {\bf r}_{i}}+\mu m^2 P_{N}\frac{\partial U}{\partial {\bf r}_{i}}\right ).
\end{eqnarray}
The stationary solution of this equation is the Gibbs canonical distribution  
\begin{eqnarray}
\label{ea3}
P_{N}=\frac{1}{Z}e^{-\beta m^2 U},
\end{eqnarray}
provided that the inverse temperature $\beta=1/(k_{B}T)$ is related to the mobility and the diffusion coefficient through the Einstein relation
\begin{eqnarray}
\label{ea4}
\beta=\frac{\mu}{D_{*}}.
\end{eqnarray}
From the Fokker-Planck equation (\ref{ea2}), we can construct a
BBGKY-like hierarchy \cite{hb2}.  The exact first equation of the
hierarchy is
\begin{equation}
\label{ea5}  {\partial P_{1}\over\partial t}=
{\partial\over\partial {\bf r}_{1}}\cdot \biggl\lbrack D_{*} {\partial
P_{1}\over\partial {\bf r}_{1}}+\mu m^{2}(N-1)\int
 {\partial {u}_{12}\over\partial {\bf r}_{1}}P_{2}
d{\bf r}_{2}\biggr\rbrack.
\end{equation}
The two-body distribution function can be written as
\begin{equation}
\label{ea5v}  
P_{2}({\bf r}_{1},{\bf r}_{2},t)=P_{1}({\bf r}_{1},t)P_{1}({\bf r}_{2},t)+P'_{2}({\bf r}_{1},{\bf r}_{2},t),
\end{equation}
where $P'_{2}({\bf r}_{1},{\bf r}_{2},t)$ is the correlation function.
Let us consider a weak long-range potential of interaction $u({\bf
r}_{12})=k\tilde{u}({\bf r}_{12})$ in a proper thermodynamic limit
$N\rightarrow +\infty$ in such a way that $k\sim 1/N$ and $V\sim 1$.
Using scaling arguments, it can be shown that $P_{2}'=O(1/N)$ except
close to a critical point (see \cite{hb1,assise} for details).
Therefore, if we are far from a critical point and if $N$ is
sufficiently large, we can make the mean field approximation
$P_{2}({\bf r}_{1},{\bf r}_{2},t)\simeq P_{1}({\bf r}_{1},t)
P_{1}({\bf r}_{2},t)$ and we obtain
\begin{eqnarray}
\label{ea6}
\frac{\partial P_{1}}{\partial t}=
{\partial\over\partial {\bf r}_{1}}\cdot \biggl\lbrack D_{*} {\partial
P_{1}\over\partial {\bf r}_{1}}+\mu m^{2}NP_{1}({\bf r},t)\int
 {\partial {u}_{12}\over\partial {\bf r}_{1}}P_{1}({\bf r}_{2},t)
d{\bf r}_{2}\biggr\rbrack.
\end{eqnarray}
Introducing the smooth density distribution $\rho({\bf r},t)=NmP_{1}({\bf r},t)$, this equation can be rewritten as 
\begin{eqnarray}
\label{ea7}
\frac{\partial\rho}{\partial t}({\bf r},t)=D_{*}\Delta\rho({\bf r},t)
+\mu m\nabla \cdot \left (\rho({\bf r},t)\nabla\int  \rho({\bf r}',t)u({\bf r}-{\bf r}')d{\bf r}'\right ).
\end{eqnarray}
It can be put in the form of a mean field Smoluchowski equation
\begin{eqnarray}
\label{ea8}
\frac{\partial\rho}{\partial t}=D_{*}\Delta\rho
+\mu m \nabla\cdot (\rho\nabla\Phi),
\end{eqnarray}
where $\Phi({\bf r},t)$ is a smooth potential produced by the
particles themselves
\begin{eqnarray}
\label{ea9}
\Phi({\bf r},t)= \int  \rho({\bf r}',t)u({\bf r}-{\bf r}')\, d{\bf r}'.
\end{eqnarray}
If we introduce the  mean field Boltzmann free energy functional 
\begin{eqnarray}
\label{ea12}
F=E-TS=\frac{1}{2}\int \rho\Phi \, d{\bf r}+k_{B}T\int \frac{\rho}{m}\ln \frac{\rho}{m}\, d{\bf r},
\end{eqnarray}
we can rewrite the mean field Smoluchowski equation (\ref{ea8}) in the form
\begin{eqnarray}
\label{ea13}
\frac{\partial\rho}{\partial t}=\nabla\cdot \left\lbrack \frac{1}{\xi}\rho({\bf r},t)\nabla\frac{\delta F}{\delta\rho}\right\rbrack.
\end{eqnarray}
This equation satisfies an $H$-theorem appropriate to the canonical ensemble
\begin{eqnarray}
\label{ea13a}
\dot F=\int \frac{\delta F}{\delta\rho}\frac{\partial\rho}{\partial t}d{\bf r}=\int \frac{\delta F}{\delta\rho}\nabla\cdot \left\lbrack \frac{1}{\xi}\rho\nabla\frac{\delta F}{\delta\rho}\right\rbrack d{\bf r}=-\int\frac{1}{\xi}\rho \left (\nabla\frac{\delta F}{\delta\rho}\right )^{2} d{\bf r}\le 0. 
\end{eqnarray}
The steady solution of the mean field Smoluchowski equation  (\ref{ea8}) or  (\ref{ea13})  corresponds to a uniform $\mu=\delta F/\delta\rho$ leading to  the mean field Boltzmann distribution
\begin{eqnarray}
\label{ea10}
\rho=Ae^{-\beta m\Phi},
\end{eqnarray}
where $\Phi({\bf r})$ is given by Eq. (\ref{ea9}).
Finally, we note that the mean field Smoluchowski equation  (\ref{ea7})   can be written in Fourier space as
\begin{eqnarray}
\label{ea11}
\frac{\partial\hat{\rho}}{\partial t}({\bf k},t)
=-D_{*} k^2 \hat{\rho}({\bf k},t)- (2\pi)^d\mu m\int {\bf k}\cdot {\bf k}'\hat{\rho}({\bf k}-{\bf k}',t)\hat{u}({\bf k}')\hat{\rho}({\bf k}',t)d{\bf k}'.
\end{eqnarray}

For weak long-range potentials of interaction, the mean field
approximation usually provides a good and useful description of the
system as a first approach \footnote{It is often advocated that
long-range potentials of interaction exhibit lack of temperedness and
stability. Some potentials, like the cosine potential in the HMF model
are well-behaved. By contrast, the gravitational potential is singular
and, strictly speaking, there is no statistical equilibrium state (no
global maximum of entropy in MCE and no global minimum of free energy
in CE).  However, there exist long-lived metastable states (local
maxima of entropy or local minima of free energy) that can be
adequately described by the mean field approximation
\cite{review}. The formation of a Dirac peak in CE, which can be
viewed as the ``equilibrium state'' of the system, can also be
described by the mean field Smoluchowski-Poisson system
\cite{post}. By contrast, the formation of binary stars in MCE
requires going beyond the mean field approximation.}. We must,
however, recall its domain of validity: (i) first, it assumes that the
number of particles in the system is large (mathematically speaking it
is valid in the limit $N\rightarrow +\infty$). Therefore, we can
expect deviations from mean field theory due to finite $N$
effects. These deviations will become manifest for sufficiently large
times (see discussion at the end of Sec. \ref{sec_cg}). (ii) Close to
a critical point $T\rightarrow T_{c}$, the correlation function
diverges (see
\cite{hb1} and Secs. \ref{sec_f} and \ref{sec_examples}). Typically, we expect a scaling of the form
$P_{2}'\sim N^{-1}(T-T_{c})^{-1}$ so that the limits $N\rightarrow
+\infty$ and $T\rightarrow T_{c}$ do not commute (see Sec. 2.7 in
\cite{cvb}). Therefore, even for large $N$, the mean field
approximation is expected to break down close to a critical point
because $P_{2}'$ is not necessarily small (the mean field
approximation is valid for $N\gg (T-T_{c})^{-1}$, which requires
larger and larger particle numbers as $T\rightarrow T_{c}$). In the two
cases (i) and (ii) mentioned above, we must take fluctuations into
account and consider stochastic kinetic equations as discussed in the
sequel.

\subsection{The exact density distribution}
\label{sec_ex}

The exact density distribution of the particles is expressed as a sum
of Dirac distributions in the form
\begin{equation}
\label{ex1}
\rho_{d}({\bf r},t)=m \sum_{i=1}^{N}\delta({\bf r}-{\bf r}_{i}(t)).
\end{equation}
It was shown by Dean \cite{dean} (see also \cite{kk,mt,arb}) that the exact density field satisfies a stochastic
equation of the form
\begin{eqnarray}
\label{ex2}
\frac{\partial\rho_d}{\partial t}({\bf r},t)=D_{*}\Delta\rho_{d}({\bf r},t)
+\mu m\nabla \cdot \left (\rho_{d}({\bf r},t)\nabla\int  \rho_{d}({\bf r}',t)u({\bf r}-{\bf r}')d{\bf r}'\right )\nonumber\\
+\nabla \cdot \left (\sqrt{2D_{*} m\rho_{d}({\bf r},t)}{\bf R}({\bf r},t)\right ),\qquad 
\end{eqnarray}
where ${\bf R}({\bf r},t)$ is a Gaussian random field such that
$\langle {\bf R}({\bf r},t)\rangle={\bf 0}$ and $\langle
R^{\alpha}({\bf r},t)R^{\beta}({\bf
r}',t')\rangle=\delta_{\alpha\beta}\delta({\bf r}-{\bf
r}')\delta(t-t')$. Note that the noise is multiplicative
\cite{dean}. Introducing the exact potential
\begin{eqnarray}
\label{ex3}
\Phi_{d}({\bf r},t)= \int  \rho_{d}({\bf r}',t)u({\bf r}-{\bf r}')\, d{\bf r}',
\end{eqnarray}
the stochastic equation (\ref{ex2}) can be rewritten as
\begin{eqnarray}
\label{ex4}
\frac{\partial\rho_d}{\partial t}=D_{*}\Delta\rho_{d}
+\mu m \nabla\cdot (\rho_{d}\nabla\Phi_{d})+\nabla\cdot (\sqrt{2D_{*}m\rho_{d}}{\bf R}).
\end{eqnarray}
For $D_{*}=0$ and $\mu\neq 0$, we get the exact deterministic
equation
\begin{eqnarray}
\label{ex5}
\frac{\partial\rho_d}{\partial t}=\mu m \nabla\cdot (\rho_{d}\nabla\Phi_{d}).
\end{eqnarray}
If we introduce the discrete Boltzmann free energy functional 
\begin{eqnarray}
\label{ex6}
F_{d}=E_{d}-TS_{d}=\frac{1}{2}\int \rho_{d}\Phi_{d} \, d{\bf r}+k_{B}T\int \frac{\rho_{d}}{m}\ln \frac{\rho_{d}}{m}\, d{\bf r},
\end{eqnarray}
we can rewrite the stochastic equation (\ref{ex4}) in the form
\begin{eqnarray}
\label{ex7}
\frac{\partial\rho_{d}}{\partial t}=\nabla\cdot \left\lbrack \frac{1}{\xi}\rho_{d}({\bf r},t)\nabla\frac{\delta F_{d}}{\delta\rho_{d}}\right\rbrack+\nabla\cdot \left (\sqrt{\frac{2k_{B}T\rho_{d}}{\xi}}{\bf R}\right ).
\end{eqnarray}
This equation can be viewed as a Langevin equation for the field
${\rho}_d({\bf r},t)$. From this equation, it can be shown that
the probability of the density distribution $W[\rho_{d},t]$ is
governed by the Fokker-Planck equation
\begin{eqnarray}
\label{ex7add}
\frac{\partial W[\rho_d,t]}{\partial t}=-\int \frac{\delta}{\delta\rho_{d}({\bf r},t)}\left\lbrace \nabla\cdot \rho_{d}({\bf r},t)\nabla\left\lbrack D_{*}\frac{\delta}{\delta\rho_{d}}+\mu\frac{\delta F_{d}}{\delta\rho_{d}}\right \rbrack W[\rho_d,t]\right\rbrace d{\bf r}.
\end{eqnarray}    
At equilibrium, we get the Boltzmann distribution  $W[\rho_{d}]\propto e^{-\beta (F_{d}[\rho_{d}]-\mu\int \rho_{d}d{\bf r})}$ \cite{kk,dean,mt,arb}. 

If we average Eq. (\ref{ex2}) over the noise, we find that the evolution of
the smooth density field $\rho({\bf r},t)=\langle\rho_{d}\rangle$ is
governed by an equation of the form
\begin{eqnarray}
\label{ex8}
\frac{\partial\rho}{\partial t}({\bf r},t)
=D_{*}\Delta\rho({\bf r},t)+\mu m\nabla \cdot \int  \langle \rho_{d}({\bf r},t)\rho_{d}({\bf r}',t)\rangle \nabla u({\bf r}-{\bf r}')d{\bf r}'.
\end{eqnarray}
Using  the identity (see Appendix \ref{sec_cf}):
\begin{eqnarray}
\label{ex9}
\langle \rho_d({\bf r},t)  \rho_{d}({\bf r}',t)\rangle=Nm^2 P_{1}({\bf r},t)\delta({\bf r}-{\bf r}')
+N(N-1)m^2 P_{2}({\bf r},{\bf r}',t),
\end{eqnarray}
and assuming that $\nabla u({\bf 0})={\bf 0}$, we find that Eq. (\ref{ex8})
coincides with the exact equation (\ref{ea5}) of the BBGKY-like
hierarchy giving the evolution of the one-body distribution
function. Furthermore, if we make the mean field approximation
$\langle \rho_d({\bf r},t)
\rho_{d}({\bf r}',t)\rangle\simeq \rho({\bf r},t) \rho({\bf r}',t)$,
we recover the mean field Smoluchowski equation (\ref{ea8}).

\subsection{The coarse-grained density distribution}
\label{sec_cg}

Equation (\ref{ea5}) (or equivalently Eq.  (\ref{ex8})) for the
ensemble averaged density field $\rho({\bf r},t)$ is a deterministic
equation since we have averaged over the noise. In contrast,
Eq. (\ref{ex2}) for the exact density field $\rho_{d}({\bf r},t)$ is a
stochastic equation taking into account fluctuations. However, it is
not very useful in practice since the field $\rho_{d}({\bf r},t)$ is a
sum of Dirac distributions, not a regular function. Therefore, it is
easier to directly solve the stochastic equations (\ref{ea1}) rather
than the equivalent Eq. (\ref{ex2}). Following \cite{arb}, we can keep
track of fluctuations while avoiding the problem of $\delta$-functions
by defining a ``coarse-grained'' density field $\overline{\rho}({\bf
r},t)$ obtained by averaging the exact density field on a
spatio-temporal window of finite resolution. The ``coarse-grained''
density field satisfies a stochastic equation of the form
\begin{eqnarray}
\label{cg1}
\frac{\partial\overline{\rho}}{\partial t}({\bf r},t)=D_{*}\Delta\overline{\rho}({\bf r},t)
+\mu m\nabla \cdot \left (\int  \overline{\rho}^{(2)}({\bf r},{\bf r}',t) \nabla u({\bf r}-{\bf r}')d{\bf r}'\right )\nonumber\\
+\nabla \cdot \left (\sqrt{2D_{*} m\overline{\rho}({\bf r},t)}{\bf R}({\bf r},t)\right ).\qquad 
\end{eqnarray}
where $\overline{\rho}^{(2)}({\bf r},{\bf r}',t)$ is a two-body
correlation function.  For a weak long-range potential of interaction
and for a sufficiently small spatio-temporal window, we propose to
make the approximation $\overline{\rho}^{(2)}({\bf r},{\bf
r}',t)\simeq
\overline{\rho}({\bf r},t)\overline{\rho}({\bf r}',t)$. In that case, we obtain
a stochastic  equation of the form
\begin{eqnarray}
\label{cg2}
\frac{\partial\overline{\rho}}{\partial t}({\bf r},t)=D_{*}\Delta\overline{\rho}({\bf r},t)
+\mu m \nabla \cdot \left (\overline{\rho}({\bf r},t)\int  \overline{\rho}({\bf r}',t) \nabla u({\bf r}-{\bf r}')d{\bf r}'\right )\nonumber\\
+\nabla \cdot \left (\sqrt{2D_{*} m\overline{\rho}({\bf r},t)}{\bf R}({\bf r},t)\right ).\qquad 
\end{eqnarray}
Introducing the smooth potential
\begin{eqnarray}
\label{cg3}
\overline{\Phi}({\bf r},t)= \int  \overline{\rho}({\bf r}',t)u({\bf r}-{\bf r}')\, d{\bf r}',
\end{eqnarray}
the stochastic  equation (\ref{cg2}) can be rewritten
\begin{eqnarray}
\label{cg4}
\frac{\partial\overline{\rho}}{\partial t}=D_{*}\Delta\overline{\rho}
+\mu m \nabla\cdot (\overline{\rho}\nabla\overline{\Phi})+\nabla\cdot (\sqrt{2D_{*}m\overline{\rho}}\ {\bf R}).
\end{eqnarray}
This will be called the {\it stochastic Smoluchowski equation} for the
smoothed-out density field $\overline{\rho}({\bf r},t)$. This 
equation is intermediate between Eqs. (\ref{ea7}) and (\ref{ex2}). It
keeps track of fluctuations while dealing with a continuous density field
instead of a sum of $\delta$-functions. This equation will be central
in the rest of the paper. We will see that it can reproduce the
equilibrium density correlation function (\ref{f26}) that was obtained in
Paper I from the equilibrium BBGKY-like hierarchy
\cite{hb1}. Therefore, it represents an improvement   with respect to
the mean field Smoluchowski equation (\ref{ea7}). We stress that this equation is physically distinct from Eq. (\ref{ex2}). In Appendix
\ref{sec_ll}, we propose an alternative derivation of Eq. (\ref{cg4})  by
using the general theory of fluctuations exposed by Landau \&
Lifshitz \cite{ll}.

If we introduce the coarse-grained Boltzmann free energy functional 
\begin{eqnarray}
\label{cg5}
F_{c.g.}=E_{c.g.}-TS_{c.g.}=\frac{1}{2}\int \overline{\rho}\overline{\Phi} \, d{\bf r}+k_{B}T\int \frac{\overline{\rho}}{m}\ln \frac{\overline{\rho}}{m}\, d{\bf r},
\end{eqnarray}
we can write the stochastic equation (\ref{cg4}) in the form
\begin{eqnarray}
\label{cg6}
\frac{\partial\overline{\rho}}{\partial t}=\nabla\cdot \left\lbrack \frac{1}{\xi}\overline{\rho}({\bf r},t)\nabla\frac{\delta F_{c.g.}}{\delta\overline{\rho}}\right\rbrack+\nabla\cdot \left (\sqrt{\frac{2k_{B}T\overline{\rho}}{\xi}}{\bf R}\right ).
\end{eqnarray}
This equation can be viewed as a Langevin equation for the field
$\overline{\rho}({\bf r},t)$. The evolution of the probability of the
density distribution $W[\overline{\rho},t]$ is governed by a
Fokker-Planck equation of the form (\ref{ex7add}) where $F_{d}$ and
$\rho_{d}$ are replaced by $F_{c.g.}$ and $\overline{\rho}$. At
equilibrium, we have $W[\overline{\rho}]\propto e^{-\beta
(F_{c.g.}[\overline{\rho}]-\mu\int \overline{\rho}d{\bf r})}$. For
$N\rightarrow +\infty$, the equilibrium distribution
$W[\overline{\rho}]$ is strongly peaked around the {\it global}
minimum of $F_{c.g.}[\overline{\rho}]$ at fixed mass. However, the
system can remain trapped in a metastable state (local minimum of
$F_{c.g.}[\overline{\rho}]$) for a very long time. Let us be more
precise.  If we ignore the noise term, Eq. (\ref{cg6}) reduces to
Eq. (\ref{ea13}). In that case, the system tends to a steady state
that is a {minimum} (global or local) of the free energy functional
$F_{c.g.}[\overline{\rho}]$ at fixed mass (maxima or saddle points of
free energy are linearly dynamically unstable with respect to mean
field Fokker-Planck equations
\cite{nfp}).  If the free energy admits several local minima, the
selection of the steady state will depend on a notion of {\it basin of
attraction}. Without noise, the system remains on a minimum of free
energy forever. Now, in the presence of noise, the fluctuations can
induce a {\it dynamical phase transition} from one minimum to the
other. Thus, accounting correctly for fluctuations is very important
when there exists metastable states. The probability of transition
scales as $e^{-\Delta F/k_{B}T}$ where $\Delta F$ is the barrier of
free energy between two minima. Therefore, in an infinite time, the
system will explore all the minima and will spend most time in the
global minimum for which $\Delta F$ is the largest. Now, for systems
with long-range interactions, the barrier of free energy $\Delta F$
scales as $N$ so that the probability of escape from a local minimum
is very small and behaves as $e^{-N}$.  Therefore, even if the global
minimum is in principle the most probable state, metastable states are
highly robust in practice since their lifetime scales like $e^{N}$. They 
are thus fully relevant for $N\gg 1$ \cite{meta}. These interesting
features (basin of attraction, dynamical phase transitions,
metastability,...)  would be interesting to study in more detail in
the case of systems with long-range interactions. Some results in this
direction have been reported in \cite{monaghan,ko,meta} in the
gravitational case.

\subsection{Generalized Cahn-Hilliard equations}
\label{sec_ch}

Let us assume that $u(|{\bf r}-{\bf r}'|)$ is a short-range potential
of interaction and that the preceding equation (\ref{cg4}) remains valid (to
simplify the notations, we drop the bar on the coarse-grained
fields). Setting ${\bf q}={\bf r}'-{\bf r}$ and writing
\begin{equation}
\label{ch1}
{\Phi}({\bf r},t)=\int u(q){\rho}({\bf r}+{\bf q},t)d{\bf q},
\end{equation}
we can Taylor expand ${\rho}({\bf r}+{\bf q},t)$ to second order in ${\bf q}$
so that
\begin{equation}
\label{ch2}
{\rho}({\bf r}+{\bf q},t)={\rho}({\bf r},t)+\sum_{i}\frac{\partial{\rho}}{\partial x_{i}}q_{i}+\frac{1}{2}\sum_{i,j}\frac{\partial^{2}{\rho}}{\partial x_{i}\partial x_{j}}q_{i}q_{j}.
\end{equation} 
Substituting this expansion in Eq. (\ref{ch1}), we obtain \cite{lemou}:
\begin{equation}
\label{ch3}
{\Phi}({\bf r},t)=-a \rho({\bf r},t)
-\frac{b}{2}\Delta\rho({\bf r},t), 
\end{equation}
with $a=-S_{d}\int_{0}^{+\infty} u(q) q^{d-1} dq$ and $b=-\frac{1}{d}
S_{d}\int_{0}^{+\infty} u(q) q^{d+1} dq$. Note that $l=(b/a)^{1/2}$
has the dimension of a length corresponding to the range of the
interaction.  Substituting Eq. (\ref{ch3}) in Eq. (\ref{cg5}), we can
put the free energy in the form
\begin{eqnarray}
\label{ch5}
F_{c.g.}\lbrack{\rho}\rbrack=\int \left\lbrack \frac{1}{2} (\nabla{\rho})^{2}+V({\rho})\right\rbrack d{\bf r},
\end{eqnarray}
where $V$ is the effective potential
\begin{equation}
\label{ch6}
V(\rho)=-\frac{a}{b}\rho^{2}+\frac{2k_{B}T}{m b} \rho\ln\rho+V_{0}.
\end{equation} 
In that case, Eq. (\ref{cg6}) can be rewritten
\begin{equation}
\label{ch7}
\frac{\partial{\rho}}{\partial t}=-A\nabla\cdot \left\lbrack {\rho}\nabla \left (\Delta{\rho}-V'({\rho})\right)\right\rbrack+\nabla\cdot \left (\sqrt{\frac{2k_{B}T{\rho}}{\xi}}{\bf R}\right ),
\end{equation} 
with $A=b/(2\xi)$. Substituting Eq. (\ref{ch6}) in Eq. (\ref{ch7}) we explicitly obtain
\begin{equation}
\label{ch8}
\xi\frac{\partial{\rho}}{\partial t}=\frac{k_{B}T}{m}\Delta{\rho}-\frac{a}{2}\Delta{\rho}^{2}-\frac{b}{2}\nabla\cdot ({\rho}\nabla(\Delta{\rho}))+\nabla\cdot \left (\sqrt{{2k_{B}T\xi{\rho}}}\, {\bf R}\right ). 
\end{equation} 
Without the noise term, the steady state of Eq. (\ref{cg6}), (\ref{ch7}) or (\ref{ch8})  corresponds to a uniform $\mu=\delta F/\delta\rho$ leading to 
\begin{equation}
\label{ch9}
\Delta{\rho}=V'({\rho})-\mu=-\frac{2a}{b}{\rho}+\frac{2k_{B}T}{mb}\ln{\rho}+\frac{2k_{B}T}{mb}-\mu.
\end{equation}
In particular, at $T=0$, Eq. (\ref{ch8}) reduces to
\begin{equation}
\label{ch10}
\xi\frac{\partial{\rho}}{\partial t}=-\frac{a}{2}\Delta{\rho}^{2}-\frac{b}{2}\nabla\cdot ({\rho}\nabla(\Delta{\rho})),
\end{equation} 
and its steady state is solution of the Helmholtz equation 
\begin{equation}
\label{ch11}
\Delta{\rho}+\frac{2a}{b}{\rho}=-\mu.
\end{equation}

Morphologically, Eq. (\ref{cg6}) with Eq. (\ref{ch5}), or equivalently
Eq. (\ref{ch7}), is similar to the stochastic Cahn-Hilliard equation
\cite{bray} for model B (conserved dynamics):
\begin{eqnarray}
\label{ch12}
\xi\frac{\partial\rho}{\partial t}=\Delta \frac{\delta F}{\delta\rho}+\sqrt{2\xi k_{B}T}\nabla\cdot {\bf R}, \qquad F\lbrack{\rho}\rbrack=\int \left\lbrack \frac{1}{2} (\nabla{\rho})^{2}+V({\rho})\right\rbrack d{\bf r},
\end{eqnarray}
or explicitly
\begin{equation}
\label{ch13}
\xi\frac{\partial\rho}{\partial t}=-\Delta (\Delta\rho-V'(\rho))+\sqrt{2\xi k_{B}T}\nabla\cdot {\bf R}.
\end{equation} 
There are, however, crucial differences between Eqs. (\ref{ch7}) and
(\ref{ch13}). First, the presence of the density $\rho({\bf r},t)$ in
the deterministic current and in the noise term. Secondly, in the usual
Cahn-Hilliard problem, the potential has a double-well shape of the
typical form $V(\rho)=(1-\rho^2)^2$ leading to a phase separation
while, in the present case, the potential (\ref{ch6}) is of a
different nature.

\subsection{Theory of fluctuations}
\label{sec_f}

Let us return to the stochastic Smoluchowski equation (\ref{cg6})
satisfied by the coarse-grained density distribution that we write in
the form (for convenience, we drop the bars on the coarse-grained
fields):
\begin{eqnarray}
\label{f1}
\frac{\partial\rho}{\partial t}
=\nabla\cdot \left\lbrack \frac{1}{\xi}\left (\frac{k_{B}T}{m}\nabla\rho+\rho\nabla\Phi\right )\right\rbrack+\nabla\cdot \left (\sqrt{\frac{2k_{B}T\rho}{\xi}}{\bf R}\right ),
\end{eqnarray}
\begin{eqnarray}
\label{f2}
\Phi({\bf r},t)= \int  \rho({\bf r}',t)u({\bf r}-{\bf r}')\, d{\bf r}'.
\end{eqnarray}
We wish to study the fluctuations of the density around an infinite
and homogeneous equilibrium distribution. To that purpose, we consider
small perturbations $\delta\rho({\bf r},t)$ and $\delta\Phi({\bf
r},t)$ around the steady state $\rho({\bf r})=\rho$, $\Phi({\bf
r})=\Phi$ with $\Phi=\rho\int u(x)d{\bf x}$. The linearized equations
for the perturbations are
\begin{eqnarray}
\label{f3}
\xi\frac{\partial\delta\rho}{\partial t}
=\frac{k_{B}T}{m}\Delta \delta\rho +\rho\Delta\delta\Phi+ \sqrt{2k_{B}T\xi\rho}\nabla\cdot {\bf R},
\end{eqnarray}
\begin{eqnarray}
\label{f4}
\delta\Phi({\bf r},t)= \int  \delta\rho({\bf r}',t)u({\bf r}-{\bf r}')\, d{\bf r}'.
\end{eqnarray}
We now decompose the perturbations in Fourier modes in the form
\begin{eqnarray}
\label{f5}
\delta\rho({\bf r},t)=\int\delta\hat{\rho}({\bf k},\omega)e^{i({\bf k}\cdot {\bf r}-\omega t)}d{\bf k}d\omega,
\end{eqnarray}
and similar expressions for $\delta\Phi({\bf r},t)$ and ${\bf R}({\bf
r},t)$. Taking the Fourier transform of Eqs. (\ref{f3}) and
(\ref{f4}), we obtain the algebraic equations
\begin{eqnarray}
\label{f6}
-i\xi \omega \delta\hat{\rho}_{k\omega}=-\frac{k_{B}T}{m}k^{2}\delta\hat{\rho}_{k\omega}-\rho k^2 \delta\hat{\Phi}_{k\omega}+\sqrt{2k_{B}T\xi\rho}\nabla\cdot \hat{\bf R}_{k\omega},
\end{eqnarray}
\begin{eqnarray}
\label{f7}
\delta\hat{\Phi}_{k\omega}=(2\pi)^{d}\hat{u}(k)\delta\hat{\rho}_{k\omega},
\end{eqnarray}
where we have used the fact that the integral in Eq. (\ref{f4}) is a
convolution. Solving for $\delta\hat{\rho}_{k\omega}$, we obtain
\begin{eqnarray}
\label{f8}
\left\lbrack \frac{k_{B}T}{m}k^{2}+(2\pi)^{d}\hat{u}(k)k^2\rho-i\xi\omega\right\rbrack\delta\hat{\rho}_{k\omega}=\sqrt{2k_{B}T\xi\rho} \ i k^{\mu}\hat{R}_{k\omega}^{\mu},
\end{eqnarray}
where the correlations of the Fourier transform of the noise are given by
\begin{eqnarray}
\label{f9}
\langle \hat{R}^{\mu}_{k\omega}\hat{R}^{\nu}_{k'\omega'}\rangle = \frac{1}{(2\pi)^{d+1}}\delta^{\mu\nu}\delta({\bf k}+{\bf k}')\delta(\omega+\omega').
\end{eqnarray}

Without noise (${\bf R}={\bf 0}$), Eq. (\ref{f8}) gives the dispersion
relation (see Paper II) associated with the mean field Smoluchowski equation
(\ref{ea8}), i.e.:
\begin{eqnarray}
\label{f10}
Z(k,\omega)\equiv \frac{k_{B}T}{m}k^{2}+(2\pi)^{d}\hat{u}(k)k^2\rho-i\xi\omega=0.
\end{eqnarray}
The function $Z(k,\omega)$ plays a role similar to the dielectric
function in plasma physics. The perturbation evolves exponentially
rapidly like $\delta\hat{\rho}_{k}(t)\propto
e^{\sigma(k)t}$ with a rate
given by $\sigma(k)=-\omega_{0}^{2}(k)/\xi$ where
\begin{eqnarray}
\label{f11}
\omega_{0}^{2}(k)\equiv \frac{k_{B}T}{m}k^{2}+(2\pi)^{d}\hat{u}(k)k^2\rho.
\end{eqnarray}
Thus, we find that the spatially homogeneous phase is stable with respect to the mean field Smoluchowski equation (\ref{ea8}) if $\omega_{0}^{2}(k)> 0$, i.e. 
\begin{eqnarray}
\label{f12}
1+(2\pi)^{d}\beta \rho m \hat{u}(k)>0,
\end{eqnarray}
for all modes $k$ and unstable (to some modes) otherwise. If $\hat{u}>0$, the homogeneous phase is always stable. Otherwise, a necessary condition of
instability is that
\begin{eqnarray}
\label{f13}
k_{B}T<k_{B}T_{c}\equiv (2\pi)^{d}\rho m|\hat{u}(k)|_{max}.
\end{eqnarray}
If this condition is fulfilled, the range of unstable wavenumbers is determined by
\begin{eqnarray}
\label{f14}
(2\pi)^{d}|\hat{u}(k)|>\frac{k_{B}T}{\rho m}.
\end{eqnarray}
Some explicit examples of potentials of interaction, and the
corresponding conditions of instability, are given in Paper I and in
\cite{jeansbio2}.

Let us now determine the correlations of the fluctuations around a
stable equilibrium homogeneous distribution in the presence of
noise. If we take the noise into account (${\bf R}\neq {\bf 0}$), the
Fourier transform of the density fluctuations is given by
\begin{eqnarray}
\label{f15}
\delta\hat{\rho}_{k\omega}=\frac{\sqrt{2k_{B}T\xi\rho} \ i k^{\mu}\hat{R}_{k\omega}^{\mu}}{Z(k,\omega)}.
\end{eqnarray}
Therefore, the correlations of the fluctuations in Fourier space are
\begin{eqnarray}
\label{f16}
\left\langle \delta\hat{\rho}_{k\omega}\delta\hat{\rho}_{k'\omega'}\right\rangle=\frac{-2k_{B}T\xi\rho k^{\mu}k^{'\nu}\langle \hat{R}_{k\omega}^{\mu}\hat{R}_{k'\omega'}^{\nu}\rangle}{Z(k,\omega)Z(k',\omega')}.
\end{eqnarray}
Using Eq. (\ref{f9}), we get 
\begin{eqnarray}
\label{f17}
\langle \delta\hat{\rho}_{k\omega}\delta\hat{\rho}_{k'\omega'}\rangle=\frac{1}{(2\pi)^{d+1}}\frac{2k_{B}T\xi\rho  k^{2}}{|Z(k,\omega)|^{2}}\delta({\bf k}+{\bf k}')\delta(\omega+\omega'),
\end{eqnarray}
or, more explicitly, 
\begin{eqnarray}
\label{f18}
\langle \delta\hat{\rho}_{k\omega}\delta\hat{\rho}_{k'\omega'}\rangle=\frac{1}{(2\pi)^{d+1}}\frac{2k_{B}T\xi\rho  k^{2}}{\left\lbrack \frac{k_{B}T}{m}k^{2}+(2\pi)^{d}\hat{u}(k)k^2\rho\right\rbrack^{2}+\xi^{2}\omega^{2}}\delta({\bf k}+{\bf k}')\delta(\omega+\omega').
\end{eqnarray}
The temporal correlation function of the Fourier components of the
density fluctuations is given by
\begin{eqnarray}
\label{f19}
\langle \delta\hat{\rho}_{k}(t)\delta\hat{\rho}_{k'}(t+\tau)\rangle=\int \langle \delta\hat{\rho}_{k\omega}\delta\hat{\rho}_{k'\omega'}\rangle e^{i\omega t}e^{i\omega'(t+\tau)}d\omega d\omega'.
\end{eqnarray}
Using Eq. (\ref{f18}), the integral on $\omega'$ is trivial and yields 
\begin{eqnarray}
\label{f20}
\langle \delta\hat{\rho}_{k}(t)\delta\hat{\rho}_{k'}(t+\tau)\rangle=\frac{1}{(2\pi)^{d+1}}2k_{B}T\xi\rho  k^{2}\delta({\bf k}+{\bf k}')\int_{-\infty}^{+\infty} d\omega\frac{e^{i\omega\tau}}{|Z(k,\omega)|^{2}}.
\end{eqnarray}
The integral on $\omega$ can be performed by using the Cauchy residue
theorem. The poles of the integrand are the zeros of the functions
$Z(k,\omega)$ and $Z(k,\omega)^{*}$, i.e. they are solutions of the
dispersion relation (\ref{f10}) and its complex conjugate. If
$\omega_{0}^{2}(k)>0$, which is required for the stability of the
homogeneous phase, the integrand has a single pole in the upper-half
plane at $\omega=i\omega_{0}^{2}(k)/\xi$ and the residue is
$e^{-\omega_{0}^{2}(k)\tau/\xi}/\lbrack
2i\xi\omega_{0}^{2}(k)\rbrack$. Therefore, we obtain
\begin{eqnarray}
\label{f21}
\langle \delta\hat{\rho}_{k}(t)\delta\hat{\rho}_{k'}(t+\tau)\rangle=\frac{1}{(2\pi)^{d}}\frac{k_{B}T\rho  k^{2}}{\omega_{0}^{2}(k)}\delta({\bf k}+{\bf k}')e^{-\omega_{0}^{2}(k)\tau/\xi},
\end{eqnarray}
or, more explicitly,
\begin{eqnarray}
\label{f22}
\langle \delta\hat{\rho}_{k}(t)\delta\hat{\rho}_{k'}(t+\tau)\rangle=\frac{1}{(2\pi)^{d}}\frac{k_{B}T\rho }{\frac{k_{B}T}{m}+(2\pi)^{d}\hat{u}(k)\rho}\delta({\bf k}+{\bf k}') e^{-\left\lbrack \frac{k_{B}T}{m}k^2+(2\pi)^{d}\hat{u}(k)k^2\rho\right\rbrack \tau/\xi}.
\end{eqnarray}
This formula is one of the most important results of this paper. We
shall come back to its physical interpretation in
Sec. \ref{sec_examples}. The equal time  correlation function  (corresponding
to $\tau=0$) is given by
\begin{eqnarray}
\label{f23}
\langle \delta\hat{\rho}_{k}\delta\hat{\rho}_{k'}\rangle=\frac{1}{(2\pi)^{d}} \frac{\rho m}{1+(2\pi)^{d}\beta \rho m\hat{u}(k)} \delta({\bf k}+{\bf k}').
\end{eqnarray}
In the absence of any interaction between the particles ($u=0$), we get
\begin{eqnarray}
\label{f24}
\langle \delta\hat{\rho}_{k}\delta\hat{\rho}_{k'}\rangle=\frac{1}{(2\pi)^{d}}\rho m {\delta({\bf k}+{\bf k}')},
\end{eqnarray}
which corresponds to the standard result \cite{ll}:
\begin{eqnarray}
\label{f24bis}
\langle (\delta\rho)^2\rangle=\frac{m\rho}{\Delta V}.
\end{eqnarray}
On the other hand, using Eq. (\ref{f23}) and the identity (see
Appendix \ref{sec_cf}):
\begin{eqnarray}
\label{f25}
\langle \delta\hat{\rho}_{k}\delta\hat{\rho}_{k'}\rangle=\frac{1}{(2\pi)^{d}}\rho m \left\lbrack 1+(2\pi)^{d}n\hat{h}({\bf k})\right\rbrack \delta({\bf k}+{\bf k}'),
\end{eqnarray}
we find that the Fourier transform of the correlation function  is 
\begin{eqnarray}
\label{f26}
\hat{h}_{eq}({\bf k})=\frac{-\beta m^2 \hat{u}({k})}{1+(2\pi)^{d}\beta n m^2 \hat{u}({k})}.
\end{eqnarray}
This is precisely the result (I-54) obtained in Paper I by analysing
the second equation of the equilibrium BBGKY-like hierarchy and neglecting
the three-body correlation function. Therefore, the stochastic Smoluchowski 
equation (\ref{f1})-(\ref{f2}) is able to reproduce the equilibrium
two-body correlation function. On the other hand, from
Eqs. (\ref{f22}) and (\ref{f25}), the Fourier transform of the
equilibrium temporal correlation function  is
\begin{eqnarray}
\label{f27}
\hat{h}({\bf k},t,t+\tau)=\hat{h}_{eq}({\bf k}) e^{-\left\lbrack \frac{k_{B}T}{m}k^2+(2\pi)^{d}\hat{u}(k)k^2\rho\right\rbrack \tau/\xi}.
\end{eqnarray}
This can be compared to the out-of-equilibrium temporal evolution of
the equal-time spatial correlation function $\hat{h}({\bf k},t)$ given
by Eq. (II-165) of Paper II. Note that the condition of stability
(\ref{f12}) is implied by Eq. (\ref{f22}), Eq. (\ref{f27}),
Eq. (II-165) and Eq. (\ref{f23}). This completes the discussion given
in Sec. 4.2 of Paper I.

\subsection{Specific examples}
\label{sec_examples}

The physical content of formula (\ref{f21}) is very
instructive. First, we note that the temporal correlation function of
the Fourier components of the density fluctuations decreases
exponentially rapidly with a decay rate
$\sigma(k)=-\omega_{0}^{2}(k)/\xi$ that coincides with the decay rate
of a perturbation of the density governed by the deterministic mean field
Smoluchowski equation (\ref{ea8}), i.e. without noise
\footnote{Similarly, in Sec. 2.9 of Paper II, we noted that, for Hamiltonian
systems with long-range interactions, the Fourier modes of the
temporal correlation function of the force decay exponentially rapidly
with a decay rate that coincides with the decay rate of a perturbation
of the distribution function governed by the Vlasov equation.}.
According to this temporal factor, or according to the mean field
theory based on the Smoluchowski equation (\ref{ea8}), the modes
satisfying the inequality $\omega_{0}^{2}(k)\ge 0$ should be
stable. For $T\le T_{c}$, the threshold of instability corresponds to
the wavenumber(s) $k=k_{m}$ where $k_{m}$ is defined by
$\omega_{0}^{2}(k_{m})=0$. Now, we note that
the amplitude of the fluctuations in formula (\ref{f21}) behaves like
$\omega_{0}^{2}(k)^{-1}$, so that it {\it diverges} as we approach the
instability threshold $k\rightarrow k_{m}$. On the other hand, if we denote by
$k_{m}^{*}$ the value of the critical wavenumber at $T=T_c$ satisfying
$1+(2\pi)^{d}\beta_{c}\rho m \hat{u}(k_{m}^{*})=0$ and if we consider
the mode $k=k_{m}^{*}$ in Eq. (\ref{f22}), we get for $T\ge T_{c}$:
\begin{eqnarray}
\label{f22b}
\langle \delta\hat{\rho}_{k}(t)\delta\hat{\rho}_{k'}(t+\tau)\rangle=\frac{1}{(2\pi)^{d}}\frac{T\rho m}{T-T_c}\delta({\bf k}+{\bf k}') e^{-k_{B}(T-T_c)(k_{m}^{*})^{2}\tau/(\xi m)}.
\end{eqnarray} 
This formula clearly shows that the correlation function diverges at
the critical point $T=T_{c}$ for the ``dangerous''mode
$k=k_{m}^{*}$. These results imply that the mean field approximation
breaks down close to the critical point (or close to the instability
threshold) and that the instability triggering the phase transition
can occur {\it sooner} than what is predicted by mean field theory
(i.e. from the stability analysis of the mean field Smoluchowski
equation). Some results in this direction have been reported in
\cite{monaghan,ko,meta} in the gravitational case.

Let us consider specific examples for illustration (we use the
notations of Paper I). For the BMF model (the Brownian version of the
HMF model) \cite{cvb}, there exists a critical temperature
$T_{c}=kM/(4\pi)$. Considering the linear dynamical stability of a
spatially homogeneous distribution with respect to the mean field
Smoluchowski equation (\ref{disp3}), we find that
$\omega_{0}^{2}(n)=Tn^2+2\pi\hat{u}_{n}\rho n^2$ where $\hat{u}_{n}$
is given by Eq. (\ref{disp5}). The modes $n\neq
\pm 1$ decay exponentially rapidly as $e^{-Tn^{2}t/\xi}$ so they are
always stable. By contrast, the modes $n=\pm 1$ evolve in time like
$e^{-(T-T_c)t/\xi}$. For $T>T_{c}$, the perturbation is damped and for
$T<T_{c}$ the perturbation has exponential growth. In that case, the
homogeneous phase is unstable to the $n=\pm 1$ modes (see Appendix
\ref{sec_disp}). According to formula (\ref{f21}), the correlation
function of the density fluctuations can be written for the stable
modes $n\neq \pm 1$:
\begin{eqnarray}
\label{exa1}
\langle\delta\hat{\rho}_{n}(t)\delta\hat{\rho}_{m}(t+\tau)\rangle=\frac{M}{4\pi^{2}n^{2}}e^{-Tn^{2}\tau/\xi}\delta_{n,-m},
\end{eqnarray}
and for the ``dangerous modes'' $n=\pm 1$:
\begin{eqnarray}
\label{exa2}
\langle\delta\hat{\rho}_{\pm 1}(t)\delta\hat{\rho}_{m}(t+\tau)\rangle=\frac{M}{4\pi^{2}}\frac{T}{T-T_{c}}e^{-(T-T_{c})\tau/\xi}\delta_{m,\mp 1}.
\end{eqnarray}
This simple toy model, where the potential of interaction is
restricted to one Fourier mode, is very interesting for pedagogical
purposes because it clearly illustrates the discussion given
above. Considering the temporal factor in Eq. (\ref{exa2}), we see
that the correlations decay for $T>T_{c}$ with the rate given by mean field
theory. However, as we approach the critical temperature from above
($T\rightarrow T_{c}^{+}$), the amplitude of the fluctuations diverges
like $(T-T_{c})^{-1}$ implying that the mean field approximation
breaks down \footnote{This implies that the limits $N\rightarrow
+\infty$ (mean field) and $T\rightarrow T_{c}$ do not commute.} and
that the phase transition should occur for $T$ {\it strictly} above
$T_{c}$. We had also reached this conclusion in \cite{cvb} from the
study of the equilibrium BBGKY-like hierarchy.

Let us now consider the attractive Yukawa potential \cite{hb1}. A
detailed stability analysis of the homogeneous phase with respect to
the mean field Smoluchowski equation (and generalizations) has been
performed in \cite{jeansbio2}. There exists a critical temperature
$k_{B}T_{c}=S_{d}G\rho m/k_{0}^{2}$ depending on the screening length
$k_{0}^{-1}$. Furthermore, in the linear regime, the perturbation
evolves exponentially rapidly as $\delta\hat{\rho}_{k}(t)\propto
e^{\sigma(k)t}$ with a rate
\begin{eqnarray}
\label{exa3re}
\sigma(k)=-\frac{\omega_{0}^{2}(k)}{\xi}=\frac{k_{B}T}{m\xi}\frac{k^{2}}{k^{2}+k_{0}^{2}}\lbrack k_{0}^{2}(T_{c}/T-1)-k^2\rbrack.
\end{eqnarray}
For $T>T_{c}$, the homogeneous phase is always stable and
for $T<T_{c}$ it is unstable to wavenumbers $k<k_{m}(T)\equiv
k_{0}(T_c/T-1)^{1/2}$. Taking into account the fluctuations, formula
(\ref{f21}) shows that the correlation function of the density fluctuations is
\begin{eqnarray}
\label{exa3}
\langle \delta\hat{\rho}_{k}(t)\delta\hat{\rho}_{k'}(t+\tau)\rangle=\frac{\rho m}{(2\pi)^{d}} \frac{k^{2}+k_{0}^{2}}{k^{2}+k_{0}^{2}(1-T_{c}/T)}\delta({\bf k}+{\bf k}')e^{-\frac{k_{B}T}{m}\frac{k^{2}}{k^{2}+k_{0}^{2}}\lbrack k^{2}+k_{0}^{2}(1-T_{c}/T)\rbrack \frac{\tau}{\xi}}.
\end{eqnarray}
Considering the mode $k=k_{m}^{*}=0$, we see that the correlation function
diverges like $(1-T_c/T)^{-1}$ as we approach the critical temperature
$T\rightarrow T_{c}^{+}$. On the other hand, for $T<T_{c}$, we see
that the amplitude diverges like $(k^{2}-k_{m}^{2})^{-1}$ as we approach
the critical wavenumber $k\rightarrow k_{m}^{+}(T)$.  This is
particularly true for the gravitational interaction ($k_0=0$) for
which Eq. (\ref{exa3}) reduces to
\begin{eqnarray}
\label{exa4}
\langle \delta\hat{\rho}_{k}(t)\delta\hat{\rho}_{k'}(t+\tau)\rangle=\frac{\rho m}{(2\pi)^{d}} \frac{k^{2}}{k^{2}-k_{J}^2}\delta({\bf k}+{\bf k}')e^{-\frac{k_{B}T}{m}(k^{2}-k_{J}^{2})\frac{\tau}{\xi}},
\end{eqnarray}
where $k_{J}=(S_{d}Gm\beta\rho)^{1/2}$ is the Jeans
wavenumber. According to the standard Jeans analysis \cite{jeans}, the
homogeneous phase is stable against perturbations with wavenumbers
$k>k_{J}$ and it becomes unstable for $k\le k_{J}$. However, the
divergence of the correlation function as $k\rightarrow k_{J}^{+}$
suggests that the gravitational instability will take place {\it
sooner}, i.e. for smaller wavelengths than the Jeans length. This
conclusion was previously reached by Monaghan \cite{monaghan} on the
basis of a hydrodynamical model of self-gravitating system
incorporating viscosity and fluctuations.

\section{The inertial model}
\label{sec_inertial}

In this section, we generalise the previous results to the case of a
stochastic model taking into account inertial effects. This
corresponds to the damped Euler equations with a long-range potential
of interaction and a stochastic forcing. This generalization allows us
to study the correlations of the fluctuations of the velocity field
and their behaviour close to the critical point. The stochastic
Smoluchowski equation (\ref{f1})-(\ref{f2}) is recovered in a strong
friction limit $\xi\rightarrow +\infty$ by neglecting the inertial
term (l.h.s.) in Eq. (\ref{dc2}). The stochastic damped Euler
equations can find applications in certain biological models of
chemotaxis where inertial effects are relevant
\cite{gamba,filbet,bio} (see also Sec. \ref{sec_chemo}).

\subsection{The density correlations}
\label{sec_dc}

We consider the stochastic damped Euler equations
\begin{eqnarray}
\label{dc1}
\frac{\partial \rho}{\partial t}+\nabla\cdot (\rho {\bf u})=0,
\end{eqnarray}
\begin{eqnarray}
\label{dc2}
\rho\left \lbrack \frac{\partial {\bf u}}{\partial t}+({\bf u}\cdot \nabla){\bf u}\right \rbrack=-\frac{k_{B}T}{m}\nabla\rho-\rho\nabla\Phi-\xi\rho {\bf u}-\sqrt{2k_{B}T\xi\rho}\, {\bf R}({\bf r},t),
\end{eqnarray}
\begin{eqnarray}
\label{dc3}
\Phi({\bf r},t)=\int u({\bf r}-{\bf r}')\rho({\bf r}',t)\, d{\bf r}',
\end{eqnarray}
where the quantities have their usual meaning. Without the stochastic
term (${\bf R}={\bf 0}$), we recover the damped Euler equations
introduced in
\cite{gen} (see also \cite{virial2,bio,jeansbio2,nfp}). With the stochastic term, we obtain a more general model
taking into account fluctuations. The form of the noise is justified
in Appendix \ref{sec_ll} using the general theory of fluctuations of
Landau \& Lifshitz \cite{ll}. For $\xi=0$, we get the usual Euler
equations and for $\xi\rightarrow +\infty$, neglecting the inertial
term (l.h.s.) in Eq. (\ref{dc2}) and substituting the resulting
expression $\xi\rho {\bf u}\simeq
-(k_{B}T/{m})\nabla\rho-\rho\nabla\Phi-\sqrt{2k_{B}T\xi\rho}{\bf R}$
in the continuity equation (\ref{dc1}), we recover the stochastic
Smoluchowski equation (\ref{f1})-(\ref{f2}). Note that Eq. (\ref{dc2})
can be written in terms of the free energy (\ref{ea12}) as
\begin{eqnarray}
\label{mm5}
\rho\left \lbrack \frac{\partial {\bf u}}{\partial t}+({\bf u}\cdot \nabla){\bf u}\right \rbrack=-\rho\nabla \frac{\delta F}{\delta\rho}-\xi\rho {\bf u}-\sqrt{2k_{B}T\xi\rho}\, {\bf R}({\bf r},t).
\end{eqnarray}

Considering small
perturbations around a uniform distribution with $\rho({\bf r})=\rho$,
$\Phi({\bf r})=\Phi$ and ${\bf u}={\bf 0}$, we find that the
linearized equations for the perturbations are
\begin{eqnarray}
\label{dc4}
\frac{\partial \delta\rho}{\partial t}+\rho \nabla\cdot {\bf u}=0,
\end{eqnarray}
\begin{eqnarray}
\label{dc5}
\rho\frac{\partial {\bf u}}{\partial t}=-\frac{k_{B}T}{m}\nabla\delta\rho-\rho\nabla\delta\Phi-\xi\rho {\bf u}-\sqrt{2k_{B}T\xi\rho}\, {\bf R}({\bf r},t),
\end{eqnarray}
\begin{eqnarray}
\label{dc6}
\delta\Phi({\bf r},t)=\int u({\bf r}-{\bf r}')\delta\rho({\bf r}',t)\, d{\bf r}'.
\end{eqnarray}
They can be combined to give
\begin{eqnarray}
\label{dc7}
\frac{\partial^{2} \delta\rho}{\partial t^2}+\xi\frac{\partial\delta\rho}{\partial t}
=\frac{k_{B}T}{m}\Delta \delta\rho +\rho\Delta\delta\Phi+ \sqrt{2k_{B}T\xi\rho}\nabla\cdot {\bf R}.
\end{eqnarray}
If we decompose the perturbations in Fourier modes of the form $e^{i({\bf k}\cdot {\bf r}-\omega t)}$, we obtain the system of algebraic equations
\begin{eqnarray}
\label{dc8}
-i\omega\delta\hat{\rho}_{k\omega}+i\rho {\bf k}\cdot \hat{\bf u}_{k\omega}=0,
\end{eqnarray}
\begin{eqnarray}
\label{dc9}
-i\omega\rho \hat{\bf u}_{k\omega}=-\frac{k_{B}T}{m}i{\bf k}\delta\hat{\rho}_{k\omega}-i{\bf k}\rho\delta\hat{\Phi}_{k\omega}-\xi\rho \hat{\bf u}_{k\omega}-\sqrt{2k_{B}T\xi\rho}\, \hat{\bf R}_{k\omega},
\end{eqnarray}
\begin{eqnarray}
\label{dc10}
\delta\hat{\Phi}_{k\omega}=(2\pi)^{d}\hat{u}(k)\delta\hat{\rho}_{k\omega}. 
\end{eqnarray}
Solving for the density perturbation, we get
\begin{eqnarray}
\label{dc11}
\left\lbrack {\frac{k_{B}T}{m}k^{2}+(2\pi)^{d}\hat{u}(k)\rho k^2-\omega(\omega+i\xi)}\right\rbrack\delta\hat{\rho}_{k\omega}={\sqrt{2k_{B}T\xi\rho} \ i k^{\mu}\hat{R}_{k\omega}^{\mu}}.
\end{eqnarray}
Without noise (${\bf R}={\bf 0}$), Eq. (\ref{dc11}) gives the
dispersion relation associated with the mean field
damped Euler equation \cite{virial2}, i.e.
\begin{eqnarray}
\label{dc12}
Z(k,\omega)\equiv \frac{k_{B}T}{m}k^{2}+(2\pi)^{d}\hat{u}(k)\rho k^2-\omega(\omega+i\xi)=0.
\end{eqnarray}
Using the definition (\ref{f11}), the dispersion relation can be rewritten
\begin{eqnarray}
\label{dc14}
\omega^{2}+i\xi\omega-\omega_{0}^{2}(k)=0,
\end{eqnarray}
with solutions
\begin{eqnarray}
\label{dc15}
\omega(k)=\frac{-i\xi\pm\sqrt{-\xi^{2}+4\omega_{0}^{2}(k)}}{2}\equiv -i\frac{\xi}{2}\pm\Omega(k).
\end{eqnarray}
The perturbation evolves in time as $e^{-i\omega(k) t}$. If
$\omega_{0}^{2}(k)<0$, the perturbation grows exponentially rapidly in
time with a rate $\lbrack
-\xi+\sqrt{\xi^{2}+4|\omega_{0}^{2}|}\rbrack/2$. If
$\omega_{0}^{2}(k)>\xi^{2}/4$, the perturbation decays exponentially
rapidly in time with a rate $-\xi/2$ and oscillates with a pulsation
$\sqrt{4\omega_{0}^{2}-\xi^{2}}/2$. If
$0<\omega_{0}^{2}(k)<\xi^{2}/4$, the perturbation decays exponentially
rapidly in time with a rate $\lbrack
-\xi+\sqrt{\xi^{2}-4\omega_{0}^{2}}\rbrack/2$ without
oscillating. Therefore, according to mean field theory, the system is
stable iff $\omega_{0}^{2}(k)>0$ for all $k$. This returns the
condition (\ref{f12}) studied in Sec. \ref{sec_f}. For $T<T_{c}$, the
onset of instability corresponds to the wavenumber(s) $k_{m}$ such that
$\omega_{0}(k_{m})=0$.

We now consider stable modes ($\omega_{0}^{2}(k)>0$) and study the
correlations of the density fluctuations in the presence of
noise. Repeating the steps of Sec.  \ref{sec_f} going from
Eq. (\ref{f15}) to Eq. (\ref{f20}), the density correlation function
can be written
\begin{eqnarray}
\label{dc16}
\left\langle \delta\hat{\rho}_{k}(t)\delta\hat{\rho}_{k'}(t+\tau)\right\rangle=\frac{1}{(2\pi)^{d+1}} 2k_{B}T\xi\rho k^{2}I(k,\tau)\delta({\bf k}+{\bf k}'),
\end{eqnarray}
where $I(k,\tau)$ is the integral
\begin{eqnarray}
\label{dc17}
I=\int_{-\infty}^{+\infty}f(\omega)e^{i\omega\tau}d\omega,\qquad f(\omega)=\frac{1}{|Z(k,\omega)|^{2}}.
\end{eqnarray}
We can evaluate the integral with the Cauchy residue theorem. The
poles of $f(\omega)$ are the zeros of the functions $Z(k,\omega)$ and
$Z(k,\omega)^{*}$, i.e. they are solutions of the dispersion relation
(\ref{dc14}) and its complex conjugate. If $\omega_{0}^{2}(k)>0$, the function
$f(\omega)$ has two simple poles in the upper-half plane at
$\omega=i\xi/2+\Omega$ and $\omega=i\xi/2-\Omega$. The residues of
$f(\omega)e^{i\omega\tau}$ at $\omega=i\xi/2+\Omega$ and at
$\omega=i\xi/2-\Omega$ are
\begin{eqnarray}
\label{dc18}
\frac{e^{-\frac{\xi}{2}\tau}e^{i\Omega\tau}}{i\xi (i\xi+2\Omega)(2\Omega)}, \qquad \frac{e^{-\frac{\xi}{2}\tau}e^{-i\Omega\tau}}{(i\xi-2\Omega)i\xi (-2\Omega)}.
\end{eqnarray}
Using the Cauchy residue theorem, and recalling that
\begin{eqnarray}
\label{dc19}
\Omega^{2}=\omega_{0}^{2}-\frac{1}{4}\xi^{2},
\end{eqnarray}
we obtain after simplification
\begin{eqnarray}
\label{dc20}
I=\frac{\pi}{\xi\omega_{0}^{2}}e^{-\frac{\xi}{2}\tau}\left\lbrack \cos(\Omega\tau)+\frac{\xi}{2\Omega}\sin(\Omega\tau)\right\rbrack,
\end{eqnarray}
where $\Omega$ can be complex. In conclusion, the density correlation
function for the inertial model is
\begin{eqnarray}
\label{dc21}
\left\langle \delta\hat{\rho}_{k}(t)\delta\hat{\rho}_{k'}(t+\tau)\right\rangle=\frac{1}{(2\pi)^{d}}\frac{k_{B}T\rho k^2}{\omega_{0}^{2}(k)}e^{-\frac{\xi}{2}\tau}\left\lbrack \cos(\Omega\tau)+\frac{\xi}{2\Omega}\sin(\Omega\tau)\right\rbrack \delta({\bf k}+{\bf k}').
\end{eqnarray}
The interpretation of this formula is similar to the one given in
Sec. \ref{sec_examples}. In particular, we see that the amplitude of
the fluctuations diverges when $k\rightarrow k_{m}$, corresponding to
$\omega_{0}^{2}(k)\rightarrow 0$. Thus, instability should set in
slightly before we reach the range of unstable wavenumbers
(\ref{f14}). In particular, the amplitude diverges at the critical
point $T\rightarrow T_{c}^{+}$ for the ``dangerous'' wavenumber
$k_{m}^{*}$. For $\xi\rightarrow 0$ (Euler), we can make the
approximation $\Omega\simeq \omega_{0}$, and we obtain
\begin{eqnarray}
\label{dc22}
\left\langle \delta\hat{\rho}_{k}(t)\delta\hat{\rho}_{k'}(t+\tau)\right\rangle=\frac{1}{(2\pi)^{d}}\frac{k_{B}T\rho k^2}{\omega_{0}^{2}(k)}e^{-\frac{\xi}{2}\tau}\cos(\omega_{0}\tau) \delta({\bf k}+{\bf k}').
\end{eqnarray}
For $\xi\rightarrow +\infty$ (Smoluchowski), we can make the
approximations
\begin{eqnarray}
\label{dc23}
\Omega\simeq \frac{\xi}{2}i\left (1-\frac{2\omega_{0}^{2}}{\xi^{2}}\right ),\quad e^{-\frac{\xi}{2}\tau} \cos(\Omega\tau)\simeq \frac{1}{2}e^{-\frac{\omega_{0}^{2}}{\xi}\tau},\quad e^{-\frac{\xi}{2}\tau} \sin(\Omega\tau)\simeq -\frac{1}{2i}e^{-\frac{\omega_{0}^{2}}{\xi}\tau},
\end{eqnarray}
and we recover the result of Eq. (\ref{f21}) obtained in the
overdamped limit.

\subsection{The velocity correlations}
\label{sec_v}

Let us now turn to the correlations of the velocity fluctuations. From
Eq. (\ref{dc9}), the Fourier components of the velocity fluctuations
satisfy the relation
\begin{eqnarray}
\label{v1}
\rho(\xi-i\omega)\hat{\bf u}_{k\omega}=-\frac{\omega_{0}^{2}}{k^2}i{\bf k}\delta\hat{\rho}_{k\omega}-\sqrt{2k_{B}T\xi\rho}\, \hat{\bf R}_{k\omega}.
\end{eqnarray}
Therefore, the  velocity correlations can be expressed as
\begin{eqnarray}
\label{v2}
\rho^2 (\xi-i\omega)(\xi-i\omega')\left\langle \hat{u}_{k\omega}^{\mu}\hat{u}_{k'\omega'}^{\nu}\right\rangle=-\frac{\omega_{0}^{4}}{k^2 k^{'2}}k^{\mu}k^{'\nu}\left\langle \delta\hat{\rho}_{k\omega}\delta\hat{\rho}_{k'\omega'}\right\rangle +2k_{B}T\xi\rho \langle \hat{R}_{k\omega}^{\mu}\hat{R}_{k'\omega'}^{\nu}\rangle\nonumber\\
+\frac{\omega_{0}^{2}}{k^2}i k^{\mu} \sqrt{2k_{B}T\xi\rho} \langle \delta\hat{\rho}_{k\omega}\hat{R}_{k'\omega'}^{\nu}\rangle+\frac{\omega_{0}^{'2}}{k^{'2}}i k^{'\nu} \sqrt{2k_{B}T\xi\rho} \langle \delta\hat{\rho}_{k'\omega'}\hat{R}_{k\omega}^{\mu}\rangle.
\end{eqnarray}
Using the relation
\begin{eqnarray}
\label{v3}
\delta\hat{\rho}_{k\omega}=\frac{i\sqrt{2k_{B}T\xi\rho}k^{\alpha}\hat{R}_{k\omega}^{\alpha}}{Z(k,\omega)}, 
\end{eqnarray}
and  Eq. (\ref{f9}), we get
\begin{eqnarray}
\label{v4}
\rho^2 (\xi^{2}+\omega^{2})\left\langle \hat{u}_{k\omega}^{\mu}\hat{u}_{k'\omega'}^{\nu}\right\rangle=-2k_{B}T\xi\rho\frac{\omega_{0}^{4}}{k^4}\frac{k^{\mu}k^{\nu}}{|Z(k,\omega)|^{2}}k^{\alpha}k^{\beta}\langle \hat{R}_{k\omega}^{\alpha}\hat{R}_{k'\omega'}^{\beta}\rangle +2k_{B}T\xi\rho \langle \hat{R}_{k\omega}^{\mu}\hat{R}_{k'\omega'}^{\nu}\rangle\nonumber\\
- 2k_{B}T\xi\rho\frac{\omega_{0}^{2}}{k^2} \frac{ k^{\mu}k^{\alpha}}{Z(k,\omega)} \langle \hat{R}_{k\omega}^{\alpha}\hat{R}_{k'\omega'}^{\nu}\rangle- 2k_{B}T\xi\rho \frac{\omega_{0}^{2}}{k^2} \frac{k^{\nu}k^{\alpha}}{Z(k,\omega)^{*}} \langle \hat{R}_{k'\omega}^{\alpha}\hat{R}_{k\omega}^{\mu}\rangle,\qquad 
\end{eqnarray}
or, more explicitly,
\begin{eqnarray}
\label{v5}
\rho^2 (\xi^{2}+\omega^{2})\left\langle \hat{u}_{k\omega}^{\mu}\hat{u}_{k'\omega'}^{\nu}\right\rangle=\frac{1}{(2\pi)^{d+1}}2k_{B}T\xi\rho \delta({\bf k}+{\bf k}')\delta(\omega+\omega')\nonumber\\
\times \left\lbrace -\frac{\omega_{0}^{4}}{k^2}\frac{k^{\mu}k^{\nu}}{|Z(k,\omega)|^{2}}+\delta^{\mu\nu} -\frac{\omega_{0}^{2}}{k^2} \frac{ k^{\mu}k^{\nu}}{Z(k,\omega)}  -\frac{\omega_{0}^{2}}{k^2} \frac{ k^{\mu}k^{\nu}}{Z(k,\omega)^{*}}\right\rbrace.
\end{eqnarray}
Finally, using the identity
\begin{eqnarray}
\label{v6}
\frac{1}{Z(k,\omega)}+\frac{1}{Z(k,\omega)^{*}}=\frac{2{\rm Re}(Z)}{|Z(k,\omega)|^{2}}=\frac{2(\omega_{0}^{2}-\omega^{2})}{|Z(k,\omega)|^{2}},
\end{eqnarray}
the foregoing relation can be rewritten
\begin{eqnarray}
\label{v7}
\left\langle \hat{u}_{k\omega}^{\mu}\hat{u}_{k'\omega'}^{\nu}\right\rangle=\frac{1}{(2\pi)^{d+1}}\frac{2k_{B}T\xi}{\rho}\frac{1}{\xi^{2}+\omega^{2}} \delta({\bf k}+{\bf k}')\delta(\omega+\omega')\nonumber\\
\times \left\lbrace -\frac{3\omega_{0}^{4}}{k^2}\frac{k^{\mu}k^{\nu}}{|Z(k,\omega)|^{2}}+\frac{2\omega_{0}^{2}\omega^{2}}{k^2}\frac{k^{\mu}k^{\nu}}{|Z(k,\omega)|^{2}}+\delta^{\mu\nu}\right\rbrace.
\end{eqnarray}
Taking the inverse Fourier transform in $\omega$-space of this
relation, we find that the temporal correlations of the velocity
fluctuations are given by
\begin{eqnarray}
\label{v8}
\left\langle \hat{u}_{k}^{\mu}(t)\hat{u}_{k'}^{\nu}(t+\tau)\right\rangle=\frac{1}{(2\pi)^{d+1}}\frac{2k_{B}T\xi}{\rho}\delta({\bf k}+{\bf k}')
\left\lbrace -{3\omega_{0}^{4}}\frac{k^{\mu}k^{\nu}}{k^2}K+2\omega_{0}^{2}\frac{k^{\mu}k^{\nu}}{k^{2}}K'+K''\delta^{\mu\nu}\right\rbrace,
\end{eqnarray}
where we have introduced the integrals
\begin{eqnarray}
\label{v9}
K=\int_{-\infty}^{+\infty}d\omega\frac{e^{i\omega\tau}}{(\xi^{2}+\omega^{2}){|Z(k,\omega)|^{2}}},\quad  K'=\int_{-\infty}^{+\infty}d\omega\frac{\omega^2 e^{i\omega\tau}}{(\xi^{2}+\omega^{2}){|Z(k,\omega)|^{2}}},
\end{eqnarray}
\begin{eqnarray}
\label{v10}
K''=\int_{-\infty}^{+\infty}d\omega\frac{e^{i\omega\tau}}{\xi^{2}+\omega^{2}}.
\end{eqnarray}
These integrals are easily calculated with the Cauchy residue theorem. After simplification, we obtain
\begin{eqnarray}
\label{v11}
\left\langle \hat{u}_{k}^{\mu}(t)\hat{u}_{k'}^{\nu}(t+\tau)\right\rangle=\frac{1}{(2\pi)^{d}}\frac{k_{B}T}{\rho}\delta({\bf k}+{\bf k}') \biggl\lbrace {\omega_{0}^{2}}\frac{k^{\mu}k^{\nu}}{k^2}\frac{1}{\omega_{0}^{2}+2\xi^{2}}e^{-\frac{\xi}{2}\tau}\biggl\lbrack\left (\frac{2\xi^{2}}{\omega_{0}^{2}}-1\right )\cos(\Omega\tau)\nonumber\\
-\frac{\xi}{\Omega} \left (\frac{7}{2}+\frac{\xi^{2}}{\omega_{0}^{2}}\right )\sin(\Omega\tau)-\left (3+\frac{2\xi^{2}}{\omega_{0}^{2}}\right ) e^{-\frac{\xi}{2}\tau}\biggr\rbrack+  e^{-{\xi}\tau}\delta^{\mu\nu}\biggr\rbrace.
\end{eqnarray}
We note that, at variance with the density correlation function (\ref{dc21}),
the velocity correlation function  does {\it not} diverge when 
$k\rightarrow k_{m}$. Indeed, for $\omega_{0}^{2}=0$, we have
\begin{eqnarray}
\label{v11b}
\left\langle \hat{u}_{k}^{\mu}(t)\hat{u}_{k'}^{\nu}(t+\tau)\right\rangle=\frac{1}{(2\pi)^{d}}\frac{k_{B}T}{\rho}e^{-\xi\tau}\delta({\bf k}+{\bf k}')\delta^{\mu\nu}.  
\end{eqnarray}
On the other hand, taking $\tau=0$ in Eq. (\ref{v11}), we obtain the equal time velocity correlation function 
\begin{eqnarray}
\label{v12}
\left\langle \hat{u}_{k}^{\mu}(t)\hat{u}_{k'}^{\nu}(t)\right\rangle=\frac{1}{(2\pi)^{d}}\frac{k_{B}T}{\rho}\delta({\bf k}+{\bf k}') \biggl ( \delta^{\mu\nu}-\frac{4\omega_{0}^{2}}{\omega_{0}^{2}+2\xi^{2}}\frac{k^{\mu}k^{\nu}}{k^2} \biggr ).
\end{eqnarray}
Contracting the indices, we get
\begin{eqnarray}
\label{v13}
\left\langle \hat{{\bf u}}_{k}(t)\cdot \hat{{\bf u}}_{k'}(t)\right\rangle=\frac{1}{(2\pi)^{d}}\frac{k_{B}T}{\rho}\delta({\bf k}+{\bf k}') \frac{2d\xi^{2}-(4-d)\omega_{0}^{2}}{\omega_{0}^{2}+2\xi^{2}}.
\end{eqnarray}
For $\xi\rightarrow 0$ (Euler), Eq. (\ref{v11}) can be simplified into
\begin{eqnarray}
\label{v14}
\left\langle \hat{u}_{k}^{\mu}(t)\hat{u}_{k'}^{\nu}(t+\tau)\right\rangle=\frac{1}{(2\pi)^{d}}\frac{k_{B}T}{\rho}\delta({\bf k}+{\bf k}') \biggl\lbrace - \frac{k^{\mu}k^{\nu}}{k^2}e^{-\frac{\xi}{2}\tau}\left\lbrack \cos(\omega_{0}\tau)+3 e^{-\frac{\xi}{2}\tau}\right\rbrack+  e^{-{\xi}\tau}\delta^{\mu\nu}\biggr\rbrace.
\end{eqnarray}
Alternatively, for $\xi\rightarrow +\infty$ (Smoluchowski), we get
\begin{eqnarray}
\label{v15}
\left\langle \hat{u}_{k}^{\mu}(t)\hat{u}_{k'}^{\nu}(t+\tau)\right\rangle=-\frac{1}{(2\pi)^{d}}\frac{3k_{B}T}{\rho}\delta({\bf k}+{\bf k}') \frac{\omega_{0}^{2}}{\xi^{2}} \frac{k^{\mu}k^{\nu}}{k^2}e^{-\frac{\omega_{0}^2}{\xi}\tau}.
\end{eqnarray}
This result can be obtained directly from the study of the stochastic Smoluchowski equation in Sec. \ref{sec_f} by defining the velocity field 
\begin{eqnarray}
\label{v16}
\xi\rho {\bf u}\equiv -\frac{k_{B}T}{m}\nabla\rho-\rho\nabla\Phi-\sqrt{2k_{B}T\xi\rho}\, {\bf R}.
\end{eqnarray}
Note that the velocity correlations in Eq. (\ref{v15}) tend to zero when $k\rightarrow k_m$ contrary to Eq. (\ref{v11b}). This shows that the limits $\xi\rightarrow +\infty$ and $k\rightarrow k_{m}$ do not commute. 

\subsection{Specific examples}
\label{sec_se}

Let us discuss specific examples by restricting ourselves, for
brevity, to the overdamped limit (\ref{v15}). For the stochastic BMF
model, we get for the stable modes $n\neq \pm 1$:
\begin{eqnarray}
\label{se1}
\left\langle \hat{u}_{n}(t)\hat{u}_{m}(t+\tau)\right\rangle=-\frac{3T^2n^2}{M\xi^{2}}\delta_{m,-n}e^{-Tn^2\tau/\xi},
\end{eqnarray}
and for the ``dangerous'' modes $n=\pm 1$: 
\begin{eqnarray}
\label{se2}
\left\langle \hat{u}_{\pm 1}(t)\hat{u}_{m}(t+\tau)\right\rangle=-\frac{3T}{M\xi^{2}}\delta_{m,\mp 1}(T-T_{c})e^{-(T-T_c)\tau/\xi}.
\end{eqnarray}
The velocity correlations tend to zero when $T\rightarrow
T_{c}^{+}$. For the attractive Yukawa potential, we get
\begin{eqnarray}
\label{se3}
\left\langle \hat{u}_{k}^{\mu}(t)\hat{u}_{k'}^{\nu}(t+\tau)\right\rangle=-\frac{3(k_{B}T)^2}{(2\pi)^{d}\xi^{2}\rho m} \frac{k^2+k_{0}^{2}(1-T_c/T)}{k^2+k_0^2} {k^{\mu}k^{\nu}}e^{-\frac{k_{B}T}{m}\frac{k^{2}}{k^{2}+k_{0}^{2}}\lbrack k^{2}+k_{0}^{2}(1-T_{c}/T)\rbrack \frac{\tau}{\xi}}\delta({\bf k}+{\bf k}').\nonumber\\
\end{eqnarray}
For $T\le T_{c}$, the amplitude tends to
zero as we approach the critical wavenumber $k\rightarrow
k_{m}^{+}(T)$.  For the gravitational
interaction ($k_0=0$), Eq. (\ref{se3}) reduces to
\begin{eqnarray}
\label{se4}
\left\langle \hat{u}_{k}^{\mu}(t)\hat{u}_{k'}^{\nu}(t+\tau)\right\rangle=-\frac{3(k_{B}T)^2}{(2\pi)^{d}\xi^{2}\rho m} (k^2-k_{J}^{2})\frac{k^{\mu}k^{\nu}}{k^2}e^{-\frac{k_{B}T}{m}(k^{2}-k_{J}^{2}) \frac{\tau}{\xi}}\delta({\bf k}+{\bf k}').
\end{eqnarray}
The amplitude goes to zero as $k\rightarrow k_{J}^{+}$.  The fact that
the velocity correlation function does not diverge when $k\rightarrow
k_{J}^{+}$ was previously observed by Monaghan \cite{monaghan} with his
hydrodynamic model.

\subsection{Stochastic model with memory}
\label{sec_mm}

The stochastic damped Euler equations (\ref{dc1})-(\ref{dc2}) can be rewritten
\begin{eqnarray}
\label{mm1}
\frac{\partial \rho}{\partial t}+\nabla\cdot (\rho {\bf u})=0,
\end{eqnarray}
\begin{eqnarray}
\label{mm2}
\frac{\partial}{\partial t}(\rho {\bf u})+\nabla (\rho {\bf u}\otimes {\bf u})=-\frac{k_{B}T}{m}\nabla\rho-\rho\nabla\Phi-\xi\rho {\bf u}-\sqrt{2k_{B}T\xi\rho}\, {\bf R}({\bf r},t).
\end{eqnarray}
If we neglect the inertial term (l.h.s.) in Eq. (\ref{mm2}) and
substitute the resulting expression for $\rho {\bf u}$ in
Eq. (\ref{mm1}), we obtain the stochastic Smoluchowski equation
(\ref{f1}). This is valid in a strong friction limit $\xi\rightarrow
+\infty$. We can obtain a more general model taking into account some
memory effects. If we neglect only the nonlinear term $\nabla (\rho {\bf u}\otimes {\bf u})$ in Eq. (\ref{mm2}), we get
\begin{eqnarray}
\label{mm3}
\frac{\partial}{\partial t}(\rho {\bf u})=-\frac{k_{B}T}{m}\nabla\rho-\rho\nabla\Phi-\xi\rho {\bf u}-\sqrt{2k_{B}T\xi\rho}\, {\bf R}({\bf r},t).
\end{eqnarray}
Taking the time derivative of Eq. (\ref{mm1}) and substituting
Eq. (\ref{mm3}) in the resulting expression, we obtain a simplified
stochastic model keeping track of memory effects
\begin{eqnarray}
\label{mm4}
\frac{\partial^{2}\rho}{\partial t^{2}}+\xi\frac{\partial\rho}{\partial t}
=\nabla\cdot  \left (\frac{k_{B}T}{m}\nabla\rho+\rho\nabla\Phi\right )+\nabla\cdot \left (\sqrt{{2k_{B}T\xi\rho}}{\bf R}\right ).
\end{eqnarray}
In terms of the free energy (\ref{ea12}), we have
\begin{eqnarray}
\label{mm6}
\frac{\partial^{2}\rho}{\partial t^{2}}+\xi\frac{\partial\rho}{\partial t}
=\nabla\cdot \left (\rho\nabla \frac{\delta F}{\delta\rho}\right ) +\nabla\cdot \left (\sqrt{{2k_{B}T\xi\rho}}{\bf R}\right ).
\end{eqnarray}
This equation, which is second order in time, is analogous to the {\it
telegraph equation} which generalizes the diffusion equation by
introducing memory effects. We note that in the linear regime $|{\bf
u}|\ll 1$ considered in Sec. \ref{sec_inertial}, the nonlinear term
$\nabla (\rho {\bf u}\otimes {\bf u})$ in Eq. (\ref{mm2}) is
negligible so that Eq. (\ref{mm4}) can be justified rigorously from
the damped Euler equations in this regime. This implies that the
theory of fluctuations that we have developed in
Sec. \ref{sec_inertial} directly applies to the stochastic equation
(\ref{mm4}). In particular, the linearization of Eq. (\ref{mm4})
around a homogeneous distribution returns Eq. (\ref{dc7}). However,
the stochastic Smoluchowski equation with memory (\ref{mm4}) may also
be relevant in the nonlinear regime as a heuristic equation. Indeed,
although we have neglected the nonlinear term $\nabla (\rho {\bf
u}\otimes {\bf u})$ in Eq. (\ref{mm2}), we have kept the full
nonlinearities in the right hand side. Therefore, Eq. (\ref{mm4}) is a
 semi-linear
model intermediate between the fully nonlinear
hydrodynamical model (\ref{mm1})-(\ref{mm2}) and the linearized
hydrodynamical model (\ref{dc7}).

Finally, we note that Eq. (\ref{mm6}) can be viewed as a form of stochastic
Cattaneo model. The deterministic Smoluchowski equation can be written
as a continuity equation $\partial_{t}\rho=-\nabla\cdot {\bf J}$ where
the current ${\bf J}=-(1/\xi)\rho\nabla\mu$ is proportional to the gradient of a
chemical potential $\mu=\delta F/\delta\rho$ \cite{nfp}. This is
similar to Fick's law for the diffusion of particles or to Fourier's
law for the diffusion of heat. In the context of heat conduction,
Cattaneo \cite{cattaneo} has proposed a modification of Fourier's law
in order to describe heat conduction with finite speed. He assumed
that the current is not instantaneously equal to the gradient
$\nabla\mu$ but relaxes to it with a time constant $1/\tau$. In the present situation, this would lead to a model of the form
\begin{eqnarray}
\label{mm7}
\frac{\partial \rho}{\partial t}+\nabla\cdot {\bf J}=0,
\end{eqnarray}
\begin{eqnarray}
\label{mm8}
\tau \frac{\partial {\bf J}}{\partial t}+{\bf J}=-\frac{1}{\xi}\rho\nabla \frac{\delta F}{\delta\rho} -\sqrt{\frac{2k_{B}T\rho}{\xi}}{\bf R},
\end{eqnarray}
where we have included the stochastic term for completeness. These
equations are equivalent to the semi-linear model formed by
Eqs. (\ref{mm1}) and (\ref{mm3}) if we set ${\bf J}=\rho {\bf u}$ and
$\tau=1/\xi$. For $\tau=0$, we recover the stochastic Smoluchowski
equation (\ref{cg6}). More generally, taking the time derivative of Eq. (\ref{mm7})
and using Eq. (\ref{mm8}), we obtain
\begin{eqnarray}
\label{mm9}
\tau\frac{\partial^{2}\rho}{\partial t^{2}}+\frac{\partial\rho}{\partial t}=\nabla\cdot \left (\frac{1}{\xi}\rho\nabla \frac{\delta F}{\delta\rho}\right ) +\nabla\cdot \left (\sqrt{\frac{2k_{B}T\rho}{\xi}}{\bf R}\right ),
\end{eqnarray}
which coincides with Eq. (\ref{mm6}) provided that we take $\tau=1/\xi$.

\subsection{Application to chemotaxis}
\label{sec_chemo}

In this section, we briefly mention the application of the preceding
results to the problem of chemotaxis in biology \cite{murray}. A more
detailed discussion is given in a specific paper \cite{chemo}
with complements and amplification. The standard Keller-Segel (KS) model
\cite{ks} of chemotaxis  can be viewed as a form of mean
field Smoluchowski equation
\cite{nfp}. It describes the diffusion of bacteria (or other
chemotactic species) in the concentration gradient of a chemical
produced by the particles themselves. As we have seen in this paper,
the correlation function diverges close to a critical point. In that
case, the mean field approximation breaks down and the fluctuations
must be taken into account.  Fluctuations also play an important role
when the particle number $N$ is small and when there exist metastable
states (local minima of free energy). In that case,
fluctuations can trigger dynamical phase transitions from one state to
the other (see Sec. \ref{sec_cg}). For these different reasons, it is
important to derive a chemotactic model going beyond the mean field
approximation and taking into account fluctuations.

We start from a microscopic model of chemotaxis where the dynamics of the
particles is governed by $N$
coupled stochastic equations of the form
\begin{eqnarray}
\label{c1}
\frac{d{\bf r}_{i}}{dt}=\chi\nabla c_{d}({\bf r}_{i}(t),t)+\sqrt{2D_{*}}{\bf R}_{i}(t),
\end{eqnarray}
\begin{eqnarray}
\label{c2}
\frac{\partial c_{d}}{\partial t}=-kc_{d}+D_{c}\Delta c_{d}+h\sum_{i=1}^{N}\delta({\bf r}-{\bf r}_{i}(t)),
\end{eqnarray}
where ${\bf r}_{i}(t)$, with $i=1,...,N$, denote the positions of the
particles and $c_{d}({\bf r},t)$ is the exact field of secreted
chemical. In these equations, $\chi$ and $D_{*}$ represent the
mobility and the diffusion coefficient of the organisms and $k$, $h$
and $D_{c}$ represent the degradation rate, the production rate and
the diffusion coefficient of the secreted chemical.  By extending
Dean's approach (and the results of Sec. \ref{sec_cg}) to the case of
chemotactic species, we obtain a stochastic Keller-Segel model of
chemotaxis:
\begin{eqnarray}
\label{c17}
\frac{\partial {\rho}}{\partial t}({\bf r},t)=D_{*}\Delta{\rho}({\bf r},t)-\chi  \nabla\cdot (\rho({\bf r},t)\nabla c({\bf r},t))
+\nabla \cdot \left (\sqrt{2D_{*}{\rho}({\bf r},t)}{\bf R}({\bf r},t)\right ),
\end{eqnarray}
\begin{eqnarray}
\label{c18}
\frac{\partial c}{\partial t}({\bf r},t)=-kc({\bf r},t)+D_{c}\Delta c({\bf r},t)+h\rho({\bf r},t),
\end{eqnarray}
generalizing the deterministic mean field Keller-Segel model. This
model fully takes into account the effect of fluctuations
\footnote{{\it Note added:} Until now, fluctuations have been ignored by
people working on chemotaxis. Therefore, our paper is the first
attempt to include fluctuations in the Keller-Segel model. However,
after submission of this paper [arXiv:0803.0263], a paper by Tailleur
\& Cates [arXiv:0803.1069] came out on a related subject. These
authors also consider the effect of fluctuations in the motion of
bacteria. They derive transport coefficients from microscopic models
but do not take into account the long-range interaction between
bacteria due to chemotaxis. Alternatively, in our approach, the
coefficients $D_*$ and $\chi$ appearing in the Langevin equations are
phenomenological coefficients but chemotaxis is fully taken into
account. Therefore, these two independent studies are complementary to
each other.}. On the other hand, there exists situations in biology
where inertial effects must be taken into account \cite{gamba}. In
that case, parabolic models like the Keller-Segel model must be
replaced by hyperbolic models similar to hydrodynamic equations
\cite{gamba,filbet,bio}. By extending the results of Sec. \ref{sec_inertial}, we obtain a hydrodynamic model of chemotaxis taking into account inertial
effects and fluctuations in the form:
\begin{eqnarray}
\label{c19}
\frac{\partial \rho}{\partial t}+\nabla\cdot (\rho {\bf u})=0,
\end{eqnarray}
\begin{eqnarray}
\label{c20}
\frac{\partial}{\partial t}(\rho {\bf u})+\nabla (\rho {\bf u}\otimes {\bf u})=-\xi D_{*}\nabla\rho+\rho\nabla c-\xi\rho {\bf u}-\sqrt{2D_{*}\xi^2\rho}\, {\bf R}({\bf r},t),
\end{eqnarray}
coupled to the field equation (\ref{c18}). In the strong friction limit
$\xi\rightarrow +\infty$ where the inertial term in Eq. (\ref{c20}) can be
neglected, it returns the stochastic KS model (\ref{c17}) with
$\chi=1/\xi$. On the other hand, if we only neglect the term $\nabla (\rho {\bf u}\otimes {\bf u})$
in Eq. (\ref{c20}) like in Sec. \ref{sec_mm}, we obtain a
stochastic equation of the form
\begin{eqnarray}
\label{c21}
\chi\frac{\partial^{2}\rho}{\partial t^{2}}+\frac{\partial\rho}{\partial t}
=\nabla\cdot (D_{*}\nabla\rho-\chi\rho\nabla c)+\nabla\cdot \left (\sqrt{2D_{*}\rho}{\bf R}\right ),
\end{eqnarray}
 It can
be viewed as a stochastic Cattaneo model of chemotaxis (or a stochastic
telegraph equation).

\section{The stochastic Kramers equation }
\label{sec_ps}

In this section, we generalize the results of
Secs. \ref{sec_ea}-\ref{sec_cg} in phase space. This is the rigorous
way to take into account inertial effects and fluctuations in the
problem. The motion of the Brownian particles is described by $N$
coupled stochastic Langevin equations of the form (see Paper I):
\begin{eqnarray}
\label{ps0}
\frac{d{\bf r}_{i}}{dt}={\bf v}_{i},
\end{eqnarray}
\begin{eqnarray}
\label{ps0b}
\frac{d{\bf v}_{i}}{dt}=-\xi {\bf v}_{i}-m\nabla_{i}U({\bf r}_{1},...,{\bf r}_{N})+\sqrt{2D}{\bf R}_{i}(t).
\end{eqnarray}
The friction coefficient $\xi$ and the diffusion
coefficient $D$ are related to each other by the Einstein relation
$\xi=D\beta m$ where $\beta=1/(k_{B}T)$ is the inverse temperature
\cite{hb1}. In the strong friction limit $\xi\rightarrow +\infty$, we
can neglect the inertial term in Eq. (\ref{ps0b}) and we obtain the overdamped
equations (\ref{ea1}) of Sec. \ref{sec_ea} with $\mu=1/(\xi m)$ and $D_{*}=D/\xi^2$.

Extending Dean's approach \cite{dean} in
phase space, we find that the exact distribution function $f_{d}({\bf
r},{\bf v},t)=m\sum_{i=1}^{N}\delta({\bf r}-{\bf r}_{i}(t))\delta({\bf
v}-{\bf v}_{i}(t))$ expressed in terms of $\delta$-functions satisfies
a stochastic equation of the form
\begin{eqnarray}
\label{ps1}
\frac{\partial f_d}{\partial t}+{\bf v}\cdot \frac{\partial f_d}{\partial {\bf r}}-\nabla\Phi_d\cdot \frac{\partial f_d}{\partial {\bf v}}=\frac{\partial}{\partial {\bf v}}\cdot \left ( D\frac{\partial f_d}{\partial {\bf v}}+\xi  f_d {\bf v}\right )+\frac{\partial}{\partial {\bf v}}\cdot (\sqrt{2Dmf_d}{\bf Q}({\bf r},{\bf v},t)),
\end{eqnarray}
where ${\bf Q}({\bf r},{\bf v},t)$ is a Gaussian random field such
that $\langle {\bf Q}({\bf r},{\bf v},t)\rangle={\bf 0}$ and $\langle
Q_{\alpha}({\bf r},{\bf v},t)Q_{\beta}({\bf r}',{\bf v}',t')\rangle\\
=\delta_{\alpha\beta}\delta({\bf r}-{\bf r}')\delta({\bf v}-{\bf
v}')\delta(t-t')$ and $\Phi_d({\bf r},t)$ is defined by Eq. (\ref{ex3}). If we average over the noise, we obtain
\begin{eqnarray}
\label{ps1av}
\frac{\partial f}{\partial t}+{\bf v}\cdot \frac{\partial f}{\partial {\bf r}}-\frac{\partial}{\partial {\bf v}}\cdot \int d{\bf r}'d{\bf v}' [\nabla u({\bf r}-{\bf r}')] \langle f_{d}({\bf r},{\bf v},t)f_{d}({\bf r}',{\bf v}',t)\rangle=\frac{\partial}{\partial {\bf v}}\cdot \left ( D\frac{\partial f}{\partial {\bf v}}+\xi  f{\bf v}\right ).
\end{eqnarray}
Using $f=NmP_{1}$ and the identity
\begin{eqnarray}
\label{idnw}
\langle f_{d}({\bf r},{\bf v},t)f_{d}({\bf r}',{\bf v}',t)\rangle=Nm^2 P_{1}({\bf r},{\bf v},t)\delta ({\bf r}-{\bf r'})\delta ({\bf v}-{\bf v'})+N(N-1)m^2 P_{2}({\bf r},{\bf v},{\bf r}',{\bf v}',t),
\end{eqnarray}
we find that Eq. (\ref{ps1av}) is equivalent to Eq. (II-139) obtained
from the BBGKY-like hierarchy.  If we implement a mean field
approximation $\langle f_{d}({\bf r},{\bf v},t)f_{d}({\bf r}',{\bf
v}',t)\rangle \simeq f({\bf r},{\bf v},t)f({\bf r}',{\bf v}',t)$, we
obtain the mean field Kramers equation \cite{hb2}:
\begin{eqnarray}
\label{ps1mf}
\frac{\partial f}{\partial t}+{\bf v}\cdot \frac{\partial f}{\partial {\bf r}}-\nabla\Phi\cdot \frac{\partial f}{\partial {\bf v}}=\frac{\partial}{\partial {\bf v}}\cdot \left ( D\frac{\partial f}{\partial {\bf v}}+\xi  f {\bf v}\right ),
\end{eqnarray}
where $\Phi({\bf r},t)$ is defined by Eq. (\ref{ea9}). Finally, we can
heuristically propose a stochastic kinetic equation for the evolution
of the coarse-grained distribution function $\overline{f}({\bf r},{\bf
v},t)$ obtained by averaging $f_{d}({\bf r},{\bf v},t)$ over a small
spatio-temporal window.  This leads to the stochastic Kramers equation
\begin{eqnarray}
\label{ps1cg}
\frac{\partial \overline{f}}{\partial t}+{\bf v}\cdot \frac{\partial \overline{f}}{\partial {\bf r}}-\nabla\overline{\Phi}\cdot \frac{\partial \overline{f}}{\partial {\bf v}}=\frac{\partial}{\partial {\bf v}}\cdot \left ( D\frac{\partial \overline{f}}{\partial {\bf v}}+\xi  \overline{f} {\bf v}\right )+\frac{\partial}{\partial {\bf v}}\cdot (\sqrt{2Dm\overline{f}}{\bf Q}({\bf r},{\bf v},t)),
\end{eqnarray}
where $\overline{\Phi}({\bf r},t)$ is defined by Eq. (\ref{cg3}).
This equation keeps track of fluctuations but applies to a continuous
distribution function instead of a sum of Dirac distributions. An
altnernative derivation of this equation is proposed in Appendix
\ref{sec_ll} using the general theory of fluctuations of Landau \&
Lifshitz \cite{ll}.

Let us now try to make the link with the parabolic and hydrodynamic
models considered in Secs. \ref{sec_overdamped} and
\ref{sec_inertial}.  Taking the hydrodynamic moments on the stochastic
Kramers equation (\ref{ps1}) and proceeding as in \cite{gen,virial2}, we
obtain
\begin{eqnarray}
\label{ps2}
\frac{\partial\rho}{\partial t}+\nabla\cdot (\rho {\bf u})=0,
\end{eqnarray}
\begin{eqnarray}
\label{ps3}
\rho\left (\frac{\partial u_{i}}{\partial t}+u_{j}\frac{\partial u_{i}}{\partial x_{j}}\right )=-\frac{\partial P_{ij}}{\partial x_{j}}-\rho\frac{\partial \Phi}{\partial x_{i}}-\xi\rho u_{i}-\int \sqrt{2Dmf}Q_{i}d{\bf v}, 
\end{eqnarray}
where $\rho({\bf r},t)=\int f d{\bf v}$ is the density, ${\bf u}({\bf r},t)=(1/\rho)\int f{\bf v}d{\bf v}$ is the local velocity, ${\bf w}={\bf v}-{\bf u}({\bf r},t)$ is the relative velocity and $P_{ij}=\int f w_{i}w_{j}d{\bf v}$ is the pressure tensor. Defining ${\bf g}({\bf r},t)=\int \sqrt{2Dmf}{\bf Q}d{\bf v}$, it is clear that ${\bf g}$ is a Gaussian noise and that its correlation function is
\begin{eqnarray}
\label{ps4}
\langle g_{i}({\bf r},t)g_{j}({\bf r}',t')\rangle=2Dm\int \sqrt{f({\bf r},{\bf v},t)f({\bf r}',{\bf v}',t')}\langle Q_{i}({\bf r},{\bf v},t)Q_{j}({\bf r}',{\bf v}',t')\rangle d{\bf v}d{\bf v}'\nonumber\\
=2Dm\delta_{ij}\delta({\bf r}-{\bf r}')\delta(t-t')\int f({\bf r},{\bf v},t) d{\bf v}=2Dm\delta_{ij}\delta({\bf r}-{\bf r}')\delta(t-t')\rho({\bf r},t). 
\end{eqnarray}
Therefore, the equation for the momentum (\ref{ps3}) can be rewritten 
\begin{eqnarray}
\label{ps5}
\rho\left (\frac{\partial u_{i}}{\partial t}+u_{j}\frac{\partial u_{i}}{\partial x_{j}}\right )=-\frac{\partial P_{ij}}{\partial x_{j}}-\rho\frac{\partial \Phi}{\partial x_{i}}-\xi\rho u_{i}-\sqrt{2Dm\rho}{R}_{i}({\bf r},t).
\end{eqnarray}
This equation is not closed since the pressure tensor depends on the
next order moment of the velocity. If, following \cite{gen,virial2},
we make a local thermodynamic equilibrium (L.T.E.) approximation
$f_{LTE}({\bf r},{\bf v},t)\simeq (\beta m/{2\pi})^{d/2}
\rho({\bf r},t) e^{-\beta m w^2/2}$ to compute the
pressure tensor, we find that $P_{ij}\simeq
(k_{B}T/m)\rho\delta_{ij}$. In that case, Eqs. (\ref{ps2}) and
(\ref{ps5}) yield the stochastic damped Euler equations
(\ref{dc1})-(\ref{dc2}). We recall, however, that there is no rigorous
justification for this local thermodynamic equilibrium
approximation. Therefore, it does not appear possible to rigorously
derive the damped Euler equations (\ref{dc1})-(\ref{dc2}) from the
Kramers equation (\ref{ps1cg}). Alternatively, if we consider the
strong friction limit $\xi\rightarrow +\infty$ for fixed $\beta$,
leading to $D=\xi/(\beta m)\rightarrow +\infty$, the first term in the
r.h.s. of Eq. (\ref{ps1}) implies that $f({\bf r},{\bf v},t)\simeq
(\beta m/{2\pi})^{d/2}
\rho({\bf r},t) e^{-\beta m v^2/2}+O(1/\xi)$, ${\bf u}=O(1/\xi)$ and $P_{ij}=(k_{B}T/m)\rho\delta_{ij}+O(1/\xi)$. To leading order in $1/\xi$, Eq. (\ref{ps5}) becomes
\begin{eqnarray}
\label{ps6}
\rho {\bf u}\simeq -\frac{1}{\xi}\left (\frac{k_{B}T}{m}\nabla\rho+\rho\nabla\Phi+\sqrt{2Dm\rho}{\bf R}({\bf r},t)\right ).
\end{eqnarray}
Inserting Eq. (\ref{ps6}) in the continuity equation (\ref{ps2}) and
defining $\mu=1/(\xi m)$ and $D_{*}=D/\xi^{2}=k_{B}T/(\xi m)$, we
obtain the stochastic Smoluchowski equation (\ref{cg4}). This equation
can thus be derived from Eq. (\ref{ps1cg}) in the limit
$\xi\rightarrow +\infty$.

\section{Conclusion}

In this paper, we have developed a theory of fluctuations for a system
of Brownian particles with weak long-range interactions. Starting from
the {\it stochastic} Smoluchowski equation (\ref{f1})-(\ref{f2}),
justified in Appendix \ref{sec_ll} from the Landau \& Lifshitz general
theory, we have obtained a simple formula (\ref{f22}) for the temporal
correlation function of the Fourier components of the density
fluctuations at equilibrium (for an infinite and homogeneous
distribution). This formula shows that the correlations decay in time
with the same damping rate as the one obtained from the study of the
normal modes of the {\it deterministic} Smoluchowski equation
(\ref{ea8}), without noise. Furthermore, the amplitude of the
correlation function diverges at the critical point $T_{c}$ (or at the
instability threshold $k=k_{m}$) leading to a failure of the mean
field approximation in that case. As a result, the limits
$N\rightarrow +\infty$ and $T\rightarrow T_{c}$ do not commute and the
instability occurs strictly before the critical point as discussed in
\cite{monaghan,ko,meta} for gravitational systems. In future works, we
shall extend this theory of fluctuations to more general
models. Indeed, the method developed in this paper can be generalized
to any type of kinetic equations including fluctuations. In
particular, the structure of formula (\ref{f20}) where $Z(k,\omega)$
is a sort of ``dielectric function'' obtained from the linearized kinetic
equation without noise, has a general scope.

\appendix

\section{Correlation functions}
\label{sec_cf}

Considering the correlation function of the exact density field (\ref{ex1}), and introducing the one and two-body distributions, we find that 
\begin{eqnarray}
\label{cf1}
\langle \rho_d({\bf r})  \rho_{d}({\bf r}')\rangle=\langle m^2 \sum_{i,j}\delta({\bf r}-{\bf r}_{i})\delta({\bf r}'-{\bf r}_{j})\rangle= \langle m^2 \sum_{i=1}^{N}\delta({\bf r}-{\bf r}_{i})\delta({\bf r}'-{\bf r})\rangle\nonumber\\
+\langle m^2 \sum_{i\neq j}\delta({\bf r}-{\bf r}_{i})\delta({\bf r}'-{\bf r}_{j})\rangle = N m^2 P_{1}({\bf r}) \delta({\bf r}-{\bf r}')
+N(N-1) m^2 P_{2}({\bf r},{\bf r}').
\end{eqnarray}
Denoting by $\rho({\bf r})=\langle \rho({\bf r})\rangle=NmP_{1}({\bf r})$ the
equilibrium averaged distribution and introducing the fluctuations $\delta\rho({\bf r})=\rho_{d}({\bf r})-\rho({\bf r})$, we get
\begin{eqnarray}
\label{cf2}
\langle \delta\rho({\bf r})\delta\rho({\bf r}')\rangle =\langle \rho_{d}({\bf r})\rho_{d}({\bf r}')\rangle -{\rho}({\bf r}){\rho}({\bf r}').
\end{eqnarray}
Starting from the identity (\ref{cf1}) and introducing the correlation
function $h({\bf r},{\bf r}')$ through the defining relation
$P_{2}({\bf r},{\bf r}')=P_{1}({\bf r})P_{1}({\bf r}')[ 1+ h({\bf r},{\bf r}')+1/N]$, we
obtain at the order $O(1/N)$:
\begin{eqnarray}
\label{cf3}
\langle \delta\rho({\bf r})\delta\rho({\bf r}')\rangle =m {\rho}({\bf r})\delta({\bf r}-{\bf r}')+{\rho}({\bf r}){\rho}({\bf r}') h({\bf r},{\bf r}').
\end{eqnarray}
For a spatially homogeneous equilibrium distribution where $\rho({\bf r})=\rho=mn$ and $h({\bf r},{\bf r}')=h(|{\bf r}-{\bf r}'|)$, the Fourier transform  of the density fluctuations is
\begin{eqnarray}
\label{cf4}
\langle\delta\hat{\rho}_{\bf k}\delta\hat{\rho}_{\bf k'}\rangle=\frac{1}{(2\pi)^{d}}\rho m \left \lbrack 1+(2\pi)^{d}n\hat{h}({\bf k})\right \rbrack \delta({\bf k}+{\bf k}').
\end{eqnarray}

The equilibrium correlation function can be obtained from the
equilibrium BBGKY-like hierarchy using a Debye-H\"uckel-type of
approximation or from field theoretical methods using the Landau
approximation. Starting from the Gibbs canonical distribution in
configuration space (I-44), which is the steady state of the $N$-body
Smoluchowski equation (\ref{ea2}), we can obtain
\cite{hb1} the equilibrium BBGKY-like hierarchy (I-45).  The first two
equations of this hierarchy are
\begin{eqnarray}
\label{tl1} {\partial P_{1}\over\partial {\bf r}_{1}}=-(N-1)\beta m^{2}\int {\partial u_{12}\over\partial {\bf r}_{1}} P_{2}\, d{\bf r}_{2},
\end{eqnarray}
\begin{eqnarray}
\label{tl2}
{\partial P_{2}\over\partial {\bf r}_{1}}=-\beta m^{2} P_{2}{\partial
u_{12}\over\partial {\bf r}_{1}}-(N-2)\beta m^{2}\int {\partial
u_{13}\over\partial {\bf r}_{1}} P_{3}\, d{\bf r}_{3}.
\end{eqnarray}
Introducing the decomposition (I-14) in Eq. (\ref{tl1}),
we first obtain
\begin{eqnarray}
\label{tl3} {\partial P_{1}\over\partial {\bf r}_{1}}=-(N-1)\beta m^{2} \int {\partial u_{12}\over\partial {\bf r}_{1}} P_{1}({\bf r}_{1})P_{1}({\bf r}_{2})d{\bf r}_{2}
-(N-1)\beta m^{2} \int {\partial u_{12}\over\partial {\bf r}_{1}} P'_{2}({\bf r}_{1},{\bf r}_{2})d{\bf r}_{2}.
\end{eqnarray}
Then, introducing the decomposition (I-14)-(I-15) in Eq. (\ref{tl2}), and using
Eq. (\ref{tl3}) to simplify some terms, we get  
\begin{eqnarray}
\label{tl4}
{\partial P_{2}'\over\partial {\bf r}_{1}}=-\beta m^{2} P_{1}({\bf r}_{1})P_{1}({\bf r}_{2}){\partial u_{12}\over\partial {\bf r}_{1}}
-\beta m^{2} P_{2}'({\bf r}_{1},{\bf r}_{2}){\partial u_{12}\over\partial {\bf r}_{1}}\nonumber\\
+\beta m^{2}\int {\partial u_{13}\over\partial {\bf r}_{1}} P_{1}({\bf r}_{1})P_{1}({\bf r}_{2})P_{1}({\bf r}_{3})d{\bf r}_{3}-(N-2)\beta  m^{2}\int{\partial u_{13}\over\partial {\bf r}_{1}} P_{2}'({\bf r}_{1},{\bf r}_{2}) P_{1}({\bf r}_{3})d{\bf r}_{3}\nonumber\\
+\beta m^{2}\int {\partial u_{13}\over\partial {\bf r}_{1}}P_{2}'({\bf r}_{1},{\bf r}_{3})P_{1}({\bf r}_{2})d{\bf r}_{3}- (N-2)\beta m^{2} \int {\partial u_{13}\over\partial {\bf r}_{1}} P_{2}'({\bf r}_{2},{\bf r}_{3})P_{1}({\bf r}_{1})d{\bf r}_{3}\nonumber\\
-(N-2)\beta  m^{2}\int {\partial u_{13}\over\partial {\bf r}_{1}} P_{3}'({\bf r}_{1},{\bf r}_{2},{\bf r}_{3})d{\bf r}_{3}.
\end{eqnarray}
At the order $1/N$ in the thermodynamic limit defined in Paper I where $P_1$, $\beta$, $m$, $|{\bf r}|$ are $O(1)$, $P_{2}'$, $u$ are $O(1/N)$ and $P_{3}'$ are $O(1/N^2)$, the
foregoing equations reduce to \footnote{Note that some terms of order $1/N$ were missing in Paper I because we made the approximation $N-1\simeq N$ everywhere which is incorrect.}: 
\begin{eqnarray}
\label{tl5} {\partial P_{1}\over\partial {\bf r}_{1}}=-(N-1)\beta m^{2} \int {\partial u_{12}\over\partial {\bf r}_{1}} P_{1}({\bf r}_{1})P_{1}({\bf r}_{2})d{\bf r}_{2}
-N\beta m^{2} \int {\partial u_{12}\over\partial {\bf r}_{1}} P'_{2}({\bf r}_{1},{\bf r}_{2})d{\bf r}_{2},
\end{eqnarray}
\begin{eqnarray}
\label{tl6}
{\partial P_{2}'\over\partial {\bf r}_{1}}=-\beta m^{2} P_{1}({\bf r}_{1})P_{1}({\bf r}_{2}){\partial u_{12}\over\partial {\bf r}_{1}}
+\beta m^{2}\int {\partial u_{13}\over\partial {\bf r}_{1}} P_{1}({\bf r}_{1})P_{1}({\bf r}_{2})P_{1}({\bf r}_{3})d{\bf r}_{3}\nonumber\\
-N\beta  m^{2}\int{\partial u_{13}\over\partial {\bf r}_{1}} P_{2}'({\bf r}_{1},{\bf r}_{2}) P_{1}({\bf r}_{3})d{\bf r}_{3}
- N\beta m^{2} \int {\partial u_{13}\over\partial {\bf r}_{1}} P_{2}'({\bf r}_{2},{\bf r}_{3})P_{1}({\bf r}_{1})d{\bf r}_{3}.
\end{eqnarray}
Now, introducing $P_{2}'(1,2)=P_{1}(1)P_{1}(2)\lbrack h(1,2)+\frac{1}{N}\rbrack$ in Eq. (\ref{tl6}), using Eq. (\ref{tl5}), and neglecting terms of order $O(1/N^2)$ or smaller, we find that the correlation function satisfies
\begin{eqnarray}
\label{tl7}
h({\bf r}_{1},{\bf r}_{2})=-\beta m^2 u_{12}-N\beta m^2\int u_{13}h({\bf r}_{2},{\bf r}_{3})P_{1}({\bf r}_{3})d{\bf r}_{3},
\end{eqnarray}
where $P_{1}({\bf r})$ is given by the zeroth order Eqs. (I-20) and (I-21). We
emphasize that this relation, which was not given in Paper I, is valid
for a possibly spatially {\it inhomogeneous} equilibrium state. For a
homogeneous distribution, we recover Eq. (I-51) which can be solved in
Fourier space yielding Eq. (\ref{f26}).

It is very instructive to recover these results in a different manner
using methods of field theory \cite{lebellac}. The equilibrium
probability of the density distribution governed by the stochastic
Smoluchowski equation (\ref{cg4}) is $W[\rho]=\frac{1}{Z} e^{-\beta
(F[\rho]-\mu\int {\rho}d{\bf r})}$ with $Z=\int {\cal D}\rho\  e^{-\beta
(F[\rho]-\mu\int {\rho}d{\bf r})}$ (to simplify the notations, we drop
the bars on the coarse-grained fields).  To compute the correlation
function $G({\bf r},{\bf r}')=\langle\delta\rho({\bf
r})\delta\rho({\bf r}')\rangle$, it proves convenient to introduce an
auxiliary field $\mu({\bf r})$ and write
\begin{eqnarray}
\label{tl8}
W[\rho]=\frac{1}{Z} e^{-\beta (F[\rho]-\int \mu({\bf r}) {\rho}d{\bf r})}.
\end{eqnarray}
The equilibrium corresponds to $\mu({\bf r})=\mu$. In the Landau (mean
field) approximation, the free energy $F=-k_{B}T\ln Z$ is given by
$F\simeq F[\rho]-\int \mu({\bf r}) \rho d{\bf r}$ where $\rho({\bf
r})$ is the most probable distribution of $W[\rho]$. The maximum of
$W[\rho]$ satisfies the condition $\mu({\bf r})=\delta F/\delta\rho({\bf r})$. Using the expression (\ref{cg5}) of the free
energy, we obtain
\begin{eqnarray}
\label{tl9}
\mu({\bf r})=\int u({\bf r}-{\bf r}')\rho({\bf r}')d{\bf r}'+\frac{k_{B}T}{m}\ln\rho({\bf r}). 
\end{eqnarray}
At equilibrium, taking $\mu({\bf r})=\mu$, we recover the mean field
Boltzmann distribution (\ref{ea10}). On the other hand, taking the
functional derivative of Eq. (\ref{tl9}) with respect to $\mu({\bf
r}')$ and using the fundamental identity $G({\bf r},{\bf
r}')=k_{B}T\delta\rho({\bf r})/\delta\mu({\bf r}')$ \cite{lebellac},
we find that the equilibrium correlation function is solution of
\begin{eqnarray}
\label{tl10}
\delta({\bf r}-{\bf r}')=\beta \int u({\bf r}-{\bf r}'')G({\bf r}',{\bf r}'')d{\bf r}''+\frac{1}{m\rho({\bf r})}G({\bf r},{\bf r}'). 
\end{eqnarray} 
Finally, substituting Eq. (\ref{cf3}) in Eq. (\ref{tl10}) and
simplifying some terms, we recover Eq. (\ref{tl7}). Noting that the
partition function of the $N$-body problem can be written
\begin{eqnarray}
\label{tl11}
Z=\int e^{-\beta m^2 U}d{\bf r}_{1}...d{\bf r}_{N}=\int {\cal D}\rho({\bf r})e^{S[\rho]}e^{-\beta E[\rho]}=\int {\cal D}\rho({\bf r})e^{-\beta F[\rho]},
\end{eqnarray} 
where the sum runs  over the {\it macrostates} $\rho({\bf r})$ with mass $\int \rho({\bf r})d{\bf r}=M$ and $e^{S[\rho]}$ denotes the number of {\it microstates} associated with the macrostates $\rho({\bf r})$, we see the link between the two previously exposed methods.   

\section{Application of the Landau-Lifshitz general theory of fluctuations}
\label{sec_ll}

In this Appendix, we derive the stochastic Smoluchowski equation
(\ref{f1})-(\ref{f2}) by using the general theory of fluctuations
exposed in Landau \& Lifshitz (see \cite{ll}, Chap. XVII). We write
the equation for the density in the conservative form
\begin{eqnarray}
\label{ll0}
\frac{\partial\rho}{\partial t}=-\nabla\cdot {\bf J},
\end{eqnarray}
where ${\bf J}$ is the current: 
\begin{eqnarray}
\label{ll1}
{\bf J}=-\frac{1}{\xi}\left (\frac{k_B T}{m}\nabla\rho+\rho\nabla\Phi\right )-{\bf g}({\bf r},t).
\end{eqnarray}
The first term is the deterministic Smoluchowski current (see,
e.g., \cite{hb2}) and the second term is a stochastic term that takes
into account fluctuations. The problem at hand consists in
characterizing the stochastic term ${\bf g}({\bf r},t)$. In order to
use the general theory of fluctuations \cite{ll}, we divide the fluid
volume in small elements $\Delta V$ and take the average of each
quantity in each element. The continuum limit $\Delta V\rightarrow 0$
will be performed in the final expressions. Equations (\ref{ll0}) and
(\ref{ll1}) correspond to the equations
\begin{eqnarray}
\label{ll2}
\dot{x}_{a}=-\sum_{b}\gamma_{ab}X_{b}+y_{a},
\end{eqnarray}
of the general theory \cite{ll} provided that we make the
identifications $\dot{x}_a\rightarrow -J_{\alpha}$ and
$y_{a}\rightarrow g_{\alpha}$. The $X_{a}$ can be obtained from the
expression of the rate of production of entropy. In fact, since we are
working in the canonical ensemble, the proper thermodynamical
potential is the free energy $F=E-TS$ or, equivalently, the Massieu
function $J=S-E/T$ which is the Legendre transform of the
entropy. It can be written explicitly
\begin{eqnarray}
\label{ll3}
J=-k_{B}\int \frac{\rho}{m}\ln\frac{\rho}{m} \, d{\bf
r}-\frac{1}{2T}\int \rho\Phi \, d{\bf r}.
\end{eqnarray}
Taking the time derivative of this expression, using Eq. (\ref{ll0}),
and integrating by parts, we obtain the expression
\begin{eqnarray}
\label{ll4}
\dot J=-\int \frac{1}{T\rho}\left (\frac{k_{B}T}{m}\nabla\rho+\rho\nabla\Phi\right )\cdot {\bf J} \, d{\bf r}.
\end{eqnarray}
Note that for ${\bf g}={\bf 0}$ (no noise) we recover the appropriate form of the H-theorem valid in the canonical ensemble \cite{hb2,virial2}: 
\begin{eqnarray}
\label{ll5}
\dot J=\int \frac{1}{T\rho\xi}\left (\frac{k_{B}T}{m}\nabla\rho+\rho\nabla\Phi\right )^2 \, d{\bf r}\ge 0.
\end{eqnarray}
If we replace the integral in Eq. (\ref{ll4}) by a summation on $\Delta V$, we obtain  
\begin{eqnarray}
\label{ll6}
\dot J=-\sum \frac{1}{T\rho}\left (\frac{k_{B}T}{m}\nabla\rho+\rho\nabla\Phi\right )\cdot {\bf J} \, \Delta V.
\end{eqnarray}
According to the general theory \cite{ll}, we must also have 
\begin{eqnarray}
\label{ll7}
\dot J=-k_{B}\sum_{a}X_{a}\dot{x}_{a}.
\end{eqnarray}
Comparing Eq. (\ref{ll6}) with the general expression (\ref{ll7}), we
find that the $X_a$ are given by
\begin{eqnarray}
\label{ll8}
X_{a}\rightarrow -\frac{1}{k_{B}T\rho}\left (\frac{k_{B}T}{m}\frac{\partial\rho}{\partial x_{\alpha}}+\rho\frac{\partial \Phi}{\partial x_{\alpha}}\right ) \Delta V.
\end{eqnarray}
It is now easy to find the expression of the coefficients $\gamma_{ab}$ that appear in Eq. (\ref{ll2}). Comparing Eqs. (\ref{ll1}), (\ref{ll2}) and (\ref{ll8}), we find that
\begin{eqnarray}
\label{ll9}
\gamma_{ab}=0 \quad ({\rm if} \quad a\neq b);\qquad \gamma_{aa}=\frac{k_{B}T\rho}{\xi\Delta V}.
\end{eqnarray}
Now, the general theory of fluctuations \cite{ll} gives
\begin{eqnarray}
\label{ll10}
\langle y_{a}(t_1)y_{b}(t_2)\rangle=(\gamma_{ab}+\gamma_{ba})\delta(t_1-t_2).
\end{eqnarray}
Therefore, the correlation function of the stochastic field $g({\bf
r},t)$ satisfies
\begin{eqnarray}
\label{ll11}
\langle g_{\alpha}({\bf r},t)g_{\beta}({\bf r}',t')\rangle=0 \quad ({\rm if} \quad {\bf r}\neq {\bf r}'),
\end{eqnarray}
\begin{eqnarray}
\label{ll12}
\langle g_{\alpha}({\bf r},t)g_{\beta}({\bf r},t')\rangle=\frac{2k_{B}T\rho}{\xi\Delta V}\delta_{\alpha\beta}\delta(t-t').
\end{eqnarray}
Taking the limit $\Delta V\rightarrow 0$, we can condense the above formulae under the form
\begin{eqnarray}
\label{ll13}
\langle g_{\alpha}({\bf r},t)g_{\beta}({\bf r}',t')\rangle=\frac{2k_{B}T\rho}{\xi}\delta_{\alpha\beta}\delta(t-t')\delta({\bf r}-{\bf r}').
\end{eqnarray}
We thus recover the expression of the stochastic term appearing in
Eq. (\ref{f1}) by a method different from Dean \cite{dean}.

We can repeat the same arguments for the inertial model
(\ref{dc1})-(\ref{dc3}). We write Eq. (\ref{dc2}) in the form
\begin{eqnarray}
\label{ll14}
\rho\left \lbrack \frac{\partial {\bf u}}{\partial t}+({\bf u}\cdot \nabla){\bf u}\right \rbrack=-\frac{k_{B}T}{m}\nabla\rho-\rho\nabla\Phi-\xi\rho {\bf u}-{\bf g}'({\bf r},t),
\end{eqnarray}
where ${\bf g}'({\bf r},t)$ is the noise term to be determined.
Comparing Eq. (\ref{ll14}) with Eq. (\ref{ll2}), we have the correspondances
$\dot{x}_{a}\rightarrow -\xi \rho u_{\alpha}-g'_{\alpha}$ and $y_{a}\rightarrow
-g'_{\alpha}$. The Massieu function $J=S-E/T$ for the inertial model
is \cite{virial2}:
\begin{eqnarray}
\label{ll15}
J=-k_{B}\int \frac{\rho}{m}\ln\frac{\rho}{m} \, d{\bf r}-\frac{1}{2T}\int \rho\Phi \, d{\bf r}-\frac{1}{2T}\int \rho {\bf u}^{2}\, d{\bf r}.
\end{eqnarray}
Taking the time derivative of this expression and using
Eqs. (\ref{dc1})-(\ref{dc3}), we obtain after some elementary
calculations (see, e.g., Appendix G of  \cite{nfp}) the expression
\begin{eqnarray}
\label{ll16}
\dot J=\frac{1}{T}\int {\bf u}\cdot (\xi\rho {\bf u}+{\bf g}')\, d{\bf r}.\end{eqnarray}
For ${\bf g}'={\bf 0}$ (no noise), we recover the appropriate form of
the $H$-theorem valid in the canonical ensemble for the mean field damped Euler equation  \cite{virial2}:
\begin{eqnarray}
\dot J=\frac{1}{T}\int \xi\rho {\bf u}^{2}\, d{\bf r}\ge 0.\end{eqnarray}
The discrete expression of Eq. (\ref{ll16}) is
\begin{eqnarray}
\label{ll17}
\dot J=\frac{1}{T}\sum {\bf u}\cdot (\xi\rho {\bf u}+{\bf g}')\Delta V.
\end{eqnarray}
Comparing Eq. (\ref{ll17}) with the general expression (\ref{ll7}), we
find that the $X_a$ are given by
\begin{eqnarray}
\label{ll18}
X_{a}\rightarrow \frac{\Delta V}{k_{B}T}u_{\alpha}.
\end{eqnarray}
Then,  comparing Eqs. (\ref{ll14}), (\ref{ll2}) and (\ref{ll18}), we find that
\begin{eqnarray}
\label{ll19}
\gamma_{ab}=0 \quad ({\rm if} \quad a\neq b);\qquad \gamma_{aa}=\frac{k_{B}T\xi\rho}{\Delta V}.
\end{eqnarray}
Finally, using Eq. (\ref{ll10}), we obtain the correlation function
\begin{eqnarray}
\label{ll20}
\langle g'_{\alpha}({\bf r},t)g'_{\beta}({\bf r}',t')\rangle={2k_{B}T\xi\rho}\delta_{\alpha\beta}\delta(t-t')\delta({\bf r}-{\bf r}'),
\end{eqnarray}
which coincides with the expression given in Eq. (\ref{dc2}).

Let us finally briefly consider the kinetic model of
Sec. \ref{sec_ps}. It can be written in the form
\begin{eqnarray}
\label{ll21}
\frac{\partial f}{\partial t}+{\bf v}\cdot \frac{\partial f}{\partial {\bf r}}-\nabla\Phi\cdot \frac{\partial f}{\partial {\bf v}}=-\frac{\partial}{\partial {\bf v}}\cdot {\bf J},
\end{eqnarray}
with the current
\begin{eqnarray}
\label{ll22}
{\bf J}=-D\left (\frac{\partial f}{\partial {\bf v}}+\beta m  f {\bf v}\right )-{\bf g}({\bf r},{\bf v},t).
\end{eqnarray}
The corresponding free energy (Massieu function) $J=S-E/T$ is explicitly given by \cite{hb2}:
\begin{eqnarray}
\label{ll23}
J=-k_{B}\int \frac{f}{m}\ln \frac{f}{m}\, d{\bf r}d{\bf v}-\frac{1}{T}\int f \frac{v^{2}}{2}d{\bf r}d{\bf v}-\frac{1}{2T}\int \rho \Phi \, d{\bf r}.
\end{eqnarray}
It production rate is
\begin{eqnarray}
\label{ll24}
\dot J=-k_{B}\int \frac{1}{mf}\left (\frac{\partial f}{\partial {\bf v}}+\beta m f {\bf v}\right )\cdot {\bf J}\, d{\bf r}d{\bf v}.
\end{eqnarray}
For ${\bf g}={\bf 0}$ (no noise), we recover the appropriate form of
the $H$-theorem valid in the canonical ensemble for the mean field Kramers equation  \cite{hb2,virial2}:
\begin{eqnarray}
\label{ll25}
\dot J=k_{B}\int \frac{D}{mf}\left (\frac{\partial f}{\partial {\bf v}}+\beta m f {\bf v}\right )^{2}\, d{\bf r}d{\bf v}\ge 0.
\end{eqnarray}
On the other hand, repeating the general procedure developed previously,
we find that
\begin{eqnarray}
\label{ll26}
X_{a}\rightarrow -\frac{1}{m f}\left (\frac{\partial f}{\partial v_{\alpha}}+\beta m f v_{\alpha}\right ) \Delta {\cal V},
\end{eqnarray}
\begin{eqnarray}
\label{ll27}
\gamma_{ab}=0 \quad ({\rm if} \quad a\neq b);\qquad \gamma_{aa}=\frac{D m f}{\Delta {\cal V}},
\end{eqnarray}
where $\Delta{\cal V}$ is the elementary volume in phase
space. Passing to the limit $\Delta{\cal V}\rightarrow 0$, this leads
to a correlation function of the form
\begin{eqnarray}
\label{ll20b}
\langle g_{\alpha}({\bf r},{\bf v},t)g_{\beta}({\bf r}',{\bf v}',t')\rangle={2Dmf}\delta_{\alpha\beta}\delta(t-t')\delta({\bf r}-{\bf r}')\delta({\bf v}-{\bf v}'),
\end{eqnarray}
which coincides with the expression given in Eq. (\ref{ps1}).

\section{Dispersion relation for the inertial BMF model}
\label{sec_disp}

In this Appendix, we complement the discussion of Sec. \ref{sec_dc} by
studying the dispersion relation associated with the damped Euler
equations (\ref{dc1})-(\ref{dc3}) without noise (${\bf R}={\bf 0}$)
for the inertial BMF model
\cite{cvb}. These equations can be written as
\begin{eqnarray}
\label{disp1}
\frac{\partial\rho}{\partial t}+\frac{\partial}{\partial\theta}(\rho u)=0,
\end{eqnarray}
\begin{eqnarray}
\label{disp2}
\rho \left (\frac{\partial u}{\partial t}+u\frac{\partial u}{\partial\theta}\right )=-T\frac{\partial\rho}{\partial \theta}-\frac{k\rho}{2\pi}\int_{0}^{2\pi} \sin(\theta-\theta')\rho(\theta',t)d\theta'-\xi\rho u.
\end{eqnarray}
For $\xi=0$, they reduce to the Euler equations (see Eq. (67) of \cite{cvb}) and for $\xi\rightarrow +\infty$, we can neglect the inertial term in Eq. (\ref{disp2}) and we obtain the mean field Smoluchowski equation (see Eq. (234) of \cite{cvb}):
\begin{eqnarray}
\label{disp3}
\xi\frac{\partial\rho}{\partial t}=T\frac{\partial^{2}\rho}{\partial \theta^{2}}+\frac{k}{2\pi}\frac{\partial}{\partial\theta}\left\lbrace \rho \int_{0}^{2\pi} \sin(\theta-\theta')\rho(\theta',t)d\theta'\right\rbrace. 
\end{eqnarray}
Considering the linear dynamical stability of a homogeneous
distribution with respect to the damped Euler equations
(\ref{disp1})-(\ref{disp2}), and decomposing the perturbation in normal modes
$\delta f\sim e^{i(n\theta-\omega t)}$, the dispersion relation (\ref{dc12})
can be written 
\begin{eqnarray}
\label{disp4}
\omega(\omega+i\xi)=Tn^2+2\pi\hat{u}_{n}\rho n^2,
\end{eqnarray}
where 
\begin{eqnarray}
\label{disp5}
\hat{u}_{n}=-\frac{k}{4\pi}(\delta_{n,1}+\delta_{n,-1}).
\end{eqnarray}

\begin{figure}
\centerline{
\psfig{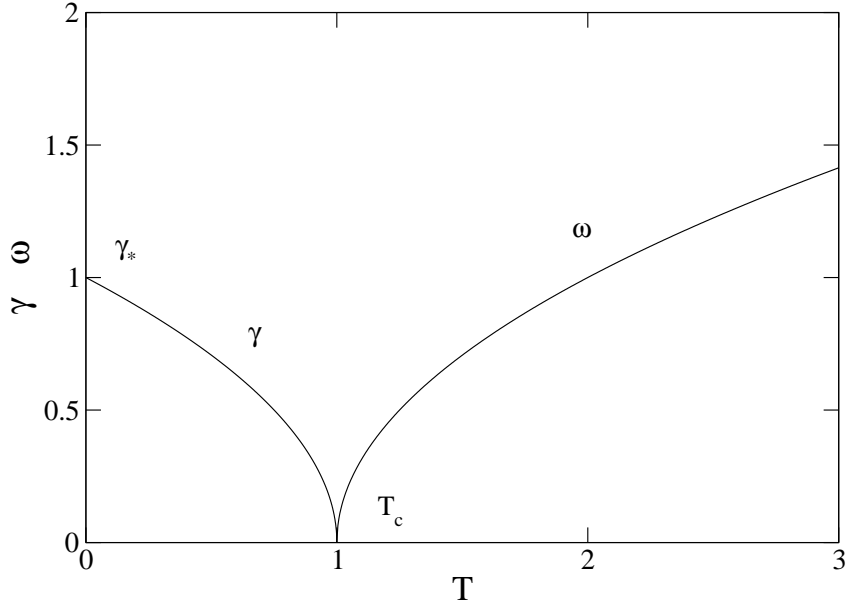}}
\caption{Evolution of the dangerous modes $n=\pm 1$ for the BMF model with $\xi=0$ (Euler). For $T>T_{c}$, the homogeneous phase is stable and the perturbation oscillates with pulsation $\omega$. For $T<T_{c}$, the homogeneous phase is unstable and the perturbation increases exponentially rapidly with a growth rate $\gamma>0$.} \label{xi0}
\end{figure}

Let us first consider the case $\xi=0$ (Euler). For $n\neq\pm 1$, the
dispersion relation becomes $\omega^2=Tn^2$ so that the perturbation
oscillates with a pulsation $\omega=\sqrt{T}|n|$. For $n=\pm 1$, the
dispersion relation becomes $\omega^2=T-T_c$ where $T_c=kM/(4\pi)$ is
the critical temperature of the BMF model \cite{cvb}. For $T>T_{c}$, the
perturbation oscillates with a pulsation $\omega=\sqrt{T-T_{c}}$ and
for $T<T_{c}$, the perturbation grows exponentially with a gowth rate
$\gamma=\sqrt{T_{c}-T}$. In that case, the homogeneous phase is
unstable (see Fig. \ref{xi0}).

\begin{figure}
\centerline{
\psfig{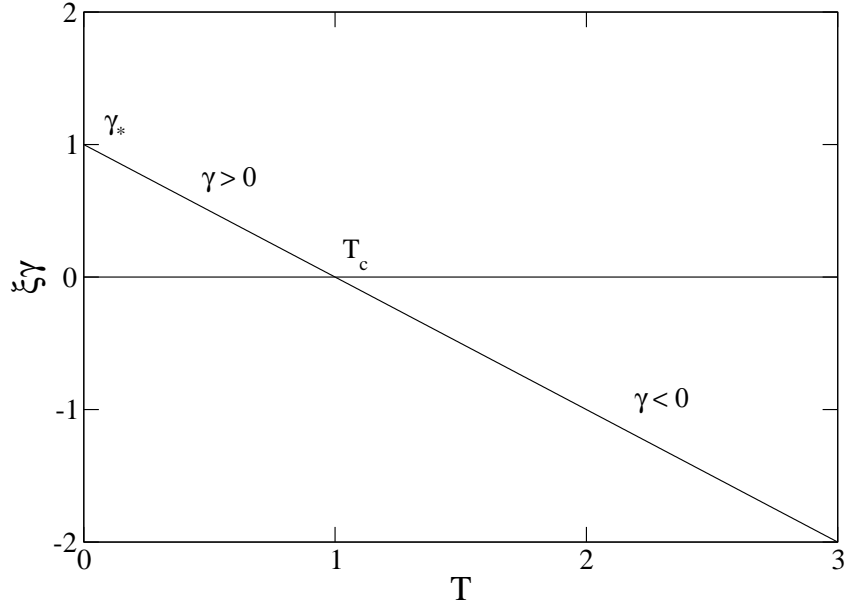}}
\caption{Evolution of the dangerous modes $n=\pm 1$ for the BMF model with $\xi\rightarrow +\infty$ (Smoluchowski). For $T>T_{c}$, the homogeneous phase is stable and the perturbation decays with a rate $\gamma<0$. For $T<T_{c}$, the homogeneous phase is unstable and the perturbation grows with a rate $\gamma>0$.} \label{xiG}
\end{figure}

\begin{figure}
\centerline{
\psfig{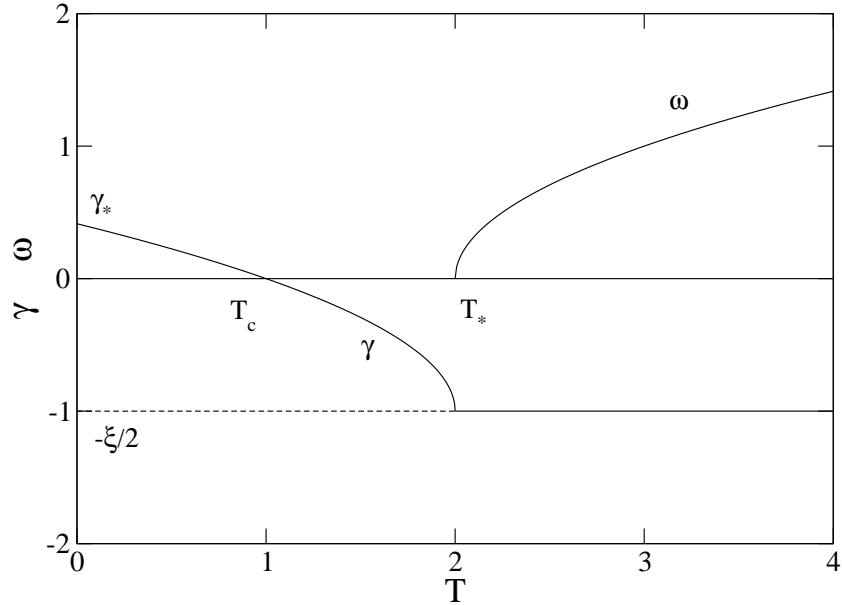}}
\caption{Evolution of the dangerous modes $n=\pm 1$ for the inertial BMF model described by the damped Euler equations. A homogeneous distribution is unstable for $T<T_{c}$ and stable for $T>T_{c}$. For $T<T_{c}$, the perturbation grows with a rate $\gamma>0$. For $T_{c}<T<T_{*}$, the perturbation decays with a rate $\gamma<0$ without oscillating. For $T>T_{*}$, the perturbation undergoes damped oscillations with a decay rate $\xi/2$ and a pulsation $\omega$. } \label{complet}
\end{figure}

Let us now consider the overdamped case $\xi\rightarrow
+\infty$ (Smoluchowski). For $n\neq\pm 1$, the dispersion
relation becomes $i\xi\omega=Tn^2$ so that the perturbation decays
exponentially with a rate $\gamma=-Tn^2/\xi$. For $n=\pm 1$, the
dispersion relation becomes $i\xi\omega=T-T_c$. For $T>T_{c}$, the
perturbation decays with a decay rate $\gamma=-(T-T_{c})/\xi$
 and for $T<T_{c}$, the
perturbation grows exponentially with a rate
$\gamma=(T_{c}-T)/\xi$. In that case, the homogeneous phase is
unstable (see Fig. \ref{xiG}).

In the general case, setting $\sigma=-i\omega$ so that the
perturbations behave as $\delta f\sim e^{\sigma t}$, the dispersion
relation (\ref{disp4}) can be rewritten
\begin{eqnarray}
\label{disp6}
\sigma^{2}+\xi\sigma+Tn^2+2\pi \hat{u}_{n}\rho n^2=0.
\end{eqnarray}
For $n\neq\pm 1$, it reduces to $\sigma^{2}+\xi\sigma+Tn^2=0$. The solutions of this equation are $\sigma_{\pm}=(-\xi\pm\sqrt{\Delta_{n}})/2$ where $\Delta_{n}=\xi^{2}-4Tn^2$. Let us introduce the  wavenumber
\begin{eqnarray}
\label{disp7}
n_{*}=\left (\frac{\xi^{2}}{4T}\right )^{1/2}.
\end{eqnarray}
For $\Delta_{n}<0$, corresponding to $n^2>n_{*}^2$, the perturbation
presents damped oscillations with a pulsation and decay rate
\begin{eqnarray}
\label{disp8}
\omega=\sqrt{T}(n^{2}-n_{*}^{2})^{1/2}, \qquad \gamma=-\xi/2.
\end{eqnarray}
For $\Delta_{n}>0$, corresponding to $n^2<n_{*}^2$ (since we have
assumed $n\neq \pm 1$, this regime is accessible iff $|n_{*}|\ge 2$
i.e. $T\le \xi^{2}/16$), the perturbation has a pure exponential
decay with a damping rate
\begin{eqnarray}
\label{disp9}
\gamma=-\frac{\xi}{2}+\sqrt{T}(n_{*}^{2}-n^{2})^{1/2}.
\end{eqnarray}
The modes $n\neq\pm 1$ are always stable, whatever the
temperature. For the ``dangerous'' modes $n=\pm 1$, the dispersion
relation becomes $\sigma^{2}+\xi\sigma+T-T_{c}=0$. The solutions are
$\sigma_{\pm}=(-\xi\pm\sqrt{\Delta_{1}})/2$ where
$\Delta_{1}=\xi^{2}-4(T-T_c)$. Let us introduce the temperature
\begin{eqnarray}
\label{disp10}
T_{*}=T_{c}+\frac{\xi^{2}}{4}.
\end{eqnarray}
For  $\Delta_{1}<0$, corresponding to  $T>T_{*}$, the perturbation undergoes damped oscillations with a pulsation and decay rate 
\begin{eqnarray}
\label{disp11}
\omega=\sqrt{T-T_{*}}, \qquad \gamma=-\xi/2.
\end{eqnarray}
For $0<\Delta_{1}<\xi^{2}$, corresponding to $T_{c}<T<T_{*}$, the perturbation has a pure exponential decay with a damping rate
\begin{eqnarray}
\label{disp12}
\gamma=-\frac{\xi}{2}+\sqrt{T_{*}-T}.
\end{eqnarray}
For $\Delta_{1}>\xi^{2}$, corresponding to $T<T_{c}$, the perturbation grows exponentially rapidly with a growth rate 
\begin{eqnarray}
\label{disp13}
\gamma=-\frac{\xi}{2}+\sqrt{T_{*}-T}.
\end{eqnarray}
Therefore, for $T<T_{c}$, the homogeneous phase is unstable to the
modes $n=\pm 1$.  The growth rate is maximum for $T=0$ with value
$\gamma_*=\gamma(0)=-{\xi}/{2}+\sqrt{T_{*}}$. The dependence of the growth
rate, damping rate and pulsation as a function of the temperature are
plotted in Fig. \ref{complet}. This figure should be compared with Fig. 1 of
\cite{jeansbio2} obtained for the gravitational potential (in this analogy, the critical temperature $T_{c}$ plays the same role as the Jeans wavenumber $k_{J}^{2}$).

It can be useful to introduce a dimensionless number
\begin{eqnarray}
\label{disp14}
F=\frac{\xi^{2}/4}{T_{c}}=\frac{\xi^{2}}{2k\rho}=\frac{\pi\xi^{2}}{kM},
\end{eqnarray}
which measures the efficiency of the friction force (a similar number
has been introduced in \cite{virial2,jeansbio2}). It can be written as
$F\sim (\xi t_{D})^{2}$ where $t_{D}\sim 1/\sqrt{k\rho}$ is a typical
dynamical time (see Sec. 2.2. of \cite{assise}).  Thus, $F$ is the ratio of the
dynamical time on the friction time $\tau\sim 1/\xi$. In terms of this
parameter, the temperature (\ref{disp10}) marking the appearance of
oscillations can be written $T_{*}=T_{c}(1+F)$.

\end{document}